\documentclass[a4paper,fleqn,usenatbib]{mnras}
\usepackage{cleveref}
\usepackage{graphicx}
\usepackage{booktabs}
\usepackage{multirow}
\usepackage{color}
\usepackage[T1]{fontenc}
\usepackage{hyperref}
\usepackage{ae,aecompl}

\newcommand{\tabitem}{~~\llap{\textbullet}~~}
\definecolor{darkgreen}{rgb}{0.0, 0.2, 0.13}

\begin{centering}\title[Machine Learning and Cosmological Simulations]{Machine Learning and Cosmological Simulations II: Hydrodynamical Simulations}
\author[Harshil M. Kamdar, Matthew J. Turk, Robert J. Brunner]{Harshil M. Kamdar$^{1,2}$\thanks{E-mail:
hkamdar2@illinois.edu}, Matthew J. Turk$^{2,4}$ and Robert J. Brunner$^{1,2,3,4,5}$ \\
$^{1}$Department of Physics, University of Illinois, Urbana, IL 61801 USA \\
$^{2}$Department of Astronomy, University of Illinois, Urbana, IL 61801 USA \\ 
$^{3}$Department of Statistics, University of Illinois, Champaign, IL 61820 USA \\
$^{4}$National Center for Supercomputing Applications, Urbana, IL 61801 USA \\
$^{5}$Beckman Institute For Advanced Science and Technology, University of Illinois, Urbana, IL, 61801 USA}
\end{centering}
\begin{document}

\date{21 October 2015}

\pagerange{\pageref{firstpage}--\pageref{lastpage}} \pubyear{2015}

\maketitle

\label{firstpage}

\begin{abstract}
We extend a machine learning (ML) framework presented previously to model galaxy
formation and evolution in a hierarchical universe using N-body + hydrodynamical
simulations. In this work, we show that ML is a promising technique to study galaxy
formation in the backdrop of a hydrodynamical simulation. We use the Illustris Simulation to train and test various sophisticated machine learning algorithms. By using
only essential dark matter halo physical properties and no merger history, our model
predicts the gas mass, stellar mass, black hole mass, star formation rate, $g-r$ color,
and stellar metallicity fairly robustly. Our results provide a unique and powerful phenomenological framework to explore the galaxy-halo connection that is built upon
a solid hydrodynamical simulation. The promising reproduction of the listed galaxy
properties demonstrably place ML as a promising and a signicantly more computationally efficient tool to study small-scale structure formation. We find that ML mimics a full-blown hydrodynamical simulation surprisingly well in a computation time
of mere minutes. The population of galaxies simulated by ML, while not numerically
identical to Illustris, is statistically robust and physically consistent with Illustris galaxies and follows the same fundamental observational constraints. Machine learning offers an intriguing and promising technique to create quick mock galaxy catalogs in the future.
\end{abstract}


\section{Introduction}
	In a $\Lambda$CDM universe, haloes grow hierarchically through mergers and gas cools  in the centers of these haloes. The evolution of collisionless dark matter particles at large scales has been studied extensively at unprecedentedly high resolutions, given the meteoric rise in computational power and the relative simplicity of these simulations \citep{springel2005cosmological, springel2005simulations, klypin2011dark, angulo2012scaling, skillman2014dark}. The formation of cosmic structure on the scale of galaxies, however, has been incredibly difficult to model \citep{baugh2006primer, somerville2014physical}; the difficulty arises primarily because baryonic physics at this scale is governed by a wide range of dissipative and/or nonlinear processes, some of which are poorly understood \citep{kang2005semianalytical, baugh2006primer, somerville2014physical}. 
    
\par
Dark matter plays an essential role in galaxy formation; broadly speaking, dark matter haloes are `cradles' of galaxy formation. It is well-established that haloes grow hierarchically through mergers and gas cools in these haloes; the evolution of galaxies, however, is dictated by a wide variety of baryonic processes that are discussed later in this paper. While baryonic physics plays a crucial role in the outcome of gaseous interactions, the story always starts with gravitational collapse. The connection between these two regimes (i.e. the galaxy-halo connection) is an important problem in modern cosmology.

\par There are two prevalent techniques used to understand galaxy formation and evolution alongside N-body dark matter simulations: semi-analytical models (hereafter, SAM) and simulations that include both hydrodynamics and gravity. The former is a post de facto technique that combines dark matter only simulations with approximate physical processes at the scale of a galaxy. For a general, exhaustive review of the motivation of SAMs and a comparison of different SAMs, the reader is referred to \citet{baugh2006primer, somerville2014physical} and \citet{knebe2015nifty}. N-body + hydrodynamical simulations (hereafter, NBHS) evolve baryonic components using fluid dynamics alongside regular dark matter evolution with subgrid models. The biggest advantage of NBHS over SAMs is the self-consistent way in which gaseous interactions are treated by the hydrodynamical codes. For a comparison of different hydrodynamical codes, the reader is referred to \citet{kim2014agora}. 

\par
In recent years, the number of hydrodynamical simulations that somewhat reproduce observed global galaxy properties has been on the rise \citep{crain2009galaxies, schaye2010physics, mccarthy2012global, puchwein2013modified, kannan2013magicc, khandai2015massiveblack, schaye2015eagle, vogelsberger2014introducing}.  The rise has been due to the rapid increase in computational power. Moreover, the subgrid models used in hydrodynamical simulations have been significantly improved for star formation \citep{springel2003cosmological, hopkins2011self}, black hole formation, and accretion \citep{sijacki2007unified, dubois2012self}. Lastly, the numerical techniques used in hydrodynamical simulations have gotten vastly more robust \citep{springel2010pur}. 

\par 
Another technique often used to create quick mock catalogs is subhalo abundance matching (SHAM) \citep{conroy2006modeling, moster2013galactic,behroozi2015simple}. SHAM assumes a monotonic relationship between the stellar mass of a galaxy and host DM halo's mass to populate a dark matter only simulation with galaxies. The similarities and the differences between SHAM and the techniques presented in this paper are presented in Section \ref{disc}.

\par 
However, it must be noted that the computational costs associated with both NBHS and SAM's are high; Illustris took a total of 19 million CPU hours to run\footnote{http://www.illustris-project.org/about/} and the largest EAGLE simulation took 4.5 million CPU hours \citep{schaye2015eagle}. Most SAMs, by construction, are meant to be significantly faster than NBHS; however, they still require an appreciable amount of computational power. For example, consider the open source GALACTICUS SAM put forth in \citet{benson2012galacticus}; a halo of mass $10^{12} M_{\odot}$ is evolved in around 2 seconds and a halo of mass $10^{15} M_{\odot}$ is evolved in around 1.25 hours. A very rough order of magnitude estimate for about 500,000 dark matter haloes, with an average evolution time of approximately 2 minutes (corresponding to about $10^{13} M_{\odot}$), implies the time taken for GALACTICUS to build merger trees to $z=0$ is $O(15,000)$ CPU hours. The inherent complexity of physical processes and the computational costs associated with a fully self-consistent treatment motivate a lot of the assumptions that SAMs make and the subgrid models employed in NBHS.

\par 
In \citet{kamdar2016machine} (hereafter referred to as K16), we explored the application of supervised machine learning (ML) techniques to the problem of galaxy formation and evolution in the backdrop of SAMs. Machine learning is a subfield of computer science that provides a platform to learn complex, non-trivial relationships in large data sets. ML has previously been applied to Astronomy with considerable success \citep{ball2006robust, ball2007robust, fiorentin2007estimation, banerji2010galaxy, ball2010data, gerdes2010arborz, kind2013tpz, xu2013first, ivezic2014statistics, ness2015cannon, kim2015hybrid, dieleman2015rotation}. As shown in K16, ML enabled the inference of some complex physical phenomena and provided a unique and powerful framework to explore the connection between the dark matter regime (large scale) and the baryonic regime (smaller, galaxy scales). 

\par
In this previous work, the Millennium simulation \citep{springel2005cosmological} along with the \citet{guo2011dwarf} SAM was used to train a few ML algorithms \citep{breiman2001random, geurts2006extremely} to predict the total stellar mass, stellar mass in the bulge, the hot gas mass, cold gas mass, and the black hole mass. The results obtained were promising for the hot gas mass, stellar mass in the bulge, and the total stellar mass with regression scores ($R^2$) of 0.99, 0.77, and 0.78 respectively; the distributions for each of these masses, the BH mass-bulge mass relation, and the stellar mass-halo mass relation were also reproduced well. The cold gas mass prediction using solely DM inputs was less robust ($R^2 = 0.39$) with severe underpredictions; it was shown that the cold gas mass prediction was not robust because of ML's inability to pick up on the time evolution of the mass cooling ODE prescribed in \citet{guo2011dwarf} by itself. The analysis was repeated with the inclusion of the cooling radius and the hot gas mass (two important ingredients in cooling ODE) over the last two snapshots and significantly better results were obtained ($R^2 = 0.82$). These promising results raise a very interesting question: can ML techniques reproduce a numerically and physically reasonable population of galaxies if trained on an N-body + hydrodynamical simulation? Furthermore, can we apply these trained ML algorithms to an N-body only simulation and reproduce a statistically reasonable population of galaxies that capture the essence of hydrodynamical simulations (i.e. essentially mimic a hydrodynamical simulation in a dark matter only simulation)? 

\par Furthermore, in K16, we also discussed \citet{neistein2010degeneracy}, where key processes were parametrized as a function of halo mass and redshift. In \citet{neistein2012hydrodynamical}, the physics from an NBHS is extracted using the same technique and are used within a SAM to explore whether a similar population of galaxies can be created. Furthermore, in \citet{dave2012analytic}, an analytical model to capture the stellar, gas, and metal content of galaxies is presented. The promising results presented in both these papers combined with results from K16 place further confidence in the approach presented in this paper. 

\par 
In this paper, we explore the feasibility of using ML to populate an hydrodynamical simulation. For our study, we use the Illustris simulation presented in \citet{vogelsberger2014introducing, vogelsberger2014properties}, and \citet{genel2014introducing}, one of the highest resolution and most ambitious N-body + hydrodynamical simulation attempted to date. The Illustris simulation has been able to reproduce a wide variety of observed galaxy properties. We extract the internal dark matter halo properties from the public database \citep{nelson2015illustris} and use the corresponding galaxy masses and star formation rate of each for our training and testing. To test the validity of our model, we make predictions at multiple epochs: $z=0,1,2,4$. A key difference between this work and K16 is the absence of merger tree history in the present work. We chose to exclude the merger history because we empirically found that their inclusion did not make a substantial difference to our results and excluding the merger history lets us train our algorithms very quickly (down from hours to minutes). 

\par
This work enables the exploration of some key questions in galaxy formation physics. How much information can be extracted from the dark matter substructures about the evolution of galaxies inside? Can similar populations of galaxies be reproduced in the case of NBHS where the physics employed is vastly more complicated? Can the approximate rules of galaxy formation be modeled by a machine to build a phenomenological model that is orders of magnitude quicker than traditional galaxy formation models? 

\par
It must be emphasized here that no baryonic processes are included in our inputs and no explicit baryonic recipes are included in our model. Our model is phenomenological; it does not seek to replace existing galaxy formation models. Instead, it serves as a powerful tool to explore the connection between dark matter haloes and their baryonic counterparts.

\par
The paper is organized as follows. In Section \ref{db}, we discuss general details about the Illustris simulation, the data extraction, the basics of machine learning, and the primary algorithms we used. In Section \ref{rd}, we present the results we obtained when the ML algorithms were applied to the Illustris data. In Section \ref{disc}, we evaluate the effectiveness of our model, point out specific deficiencies in the results we obtained, and compare the ML simulated galaxies and the Illustris galaxies. In Section \ref{cl}, we conclude the paper with an extensive summary of our findings and a discussion of future work. 

\section{Data \& Background} \label{db}

In this section, we briefly discuss the data extraction from the set of Illustris simulations and the ML algorithms that were used in our analyses. We discuss general details and briefly outline how key physical processes are handled in Illustris. Next, we briefly review the basics of ML and discuss the techniques that were employed in this work. Finally, we discuss our reasons for choosing Illustris. 

\subsection{Illustris Simulation}
What follows is only a brief overview of the Illustris simulations. For a thorough description of the physical models employed and the simulation code, the reader is referred to the following papers: \citet{springel2010pur,  vogelsberger2013model, torrey2014model, vogelsberger2014properties, vogelsberger2014introducing, genel2014introducing, sijacki2014illustris, nelson2015illustris}. The suite of Illustris simulations uses the state-of-the-art hydrodynamical code AREPO \citep{springel2010pur} to evolve resolution elements in a box of size ($106.5$ Mpc)$^3$ to $z=0$. The cosmology employed in the simulations is consistent with WMAP9: $\Omega_m = 0.2726$, $\Omega_b = 0.0456$, $\Omega_{\Lambda} = 0.7274$, $n_s = 0.963$ and $\sigma_8 = 0.809$, and the Hubble constant is  $H_0 = 70.4 $ km s$^{-1}$ Mpc$^{-1}$. Three different N-body + Hydrodynamical simulations were ran: Illustris-1 with $2 \times 1820^3$ particles, Illustris-2 with $2 \times 910^3$ particles and Illustris-3 with $2 \times 450^3$ particles. A set of N-body dark matter only simulations with the same number of particles were also ran with the same cosmology. 
\par 
Hydrodynamical equations in the Illustris simulation are solved using AREPO, a novel quasi-Lagrangian moving mesh code that uses Voronoi cells. The mesh is used to solve the Eulerian equations using finite volume techniques, while being completely Galilean invariant. The gravitational forces are calculated using the standard TreePM method \citep{xu1994new}, where short-range forces are calculated using a hierarchical octree algorithm and long-range forces are calculated using a particle mesh method. 

\par
Substructure in the Illustris simulation is identified through two different algorithms: Friends of Friends (FoF) \citep{davis1985evolution} and SUBFIND \citep{springel2001populating, dolag2009substructures}. FoF was applied to the snapshots with a linking length of 0.2 times the mean dark matter particle separation to find dark matter haloes, with at least 32 dark matter particles. The corresponding stellar, gas, and black hole elements are attached to the dark matter haloes as described in \citet{dolag2009substructures}. To find gravitationally bound structures in the simulation, a modified version of SUBFIND was used. At $z=0$, 7,713,601 FoF groups were found with more than 32 particles and 4,366,546 subhaloes were identified \citep{vogelsberger2014introducing}. 

\par
Illustris includes the treatment of several key physical processes that shape (quite literally, sometimes) galaxy formation and evolution. The simulation includes treatments of gas cooling with self-shielding corrections, star formation, black hole seeding, black hole accretion, AGN feedback (thermal quasar-mode, thermal-mechanical radio mode, radiative feedback), supernovae feedback, stellar evolution with associated chemical enrichment, stellar mass loss, and star formation feedback. The fifteen or so free parameters that are related to the modeling of the subresolution processes are finely tuned to the history of the star formation rate history and the stellar mass function at $z=0$. 
\par
Due to the limited resolution of large NBHS, a subgrid treatment of star formation is required. Like many previous simulations \citep{springel2005cosmological, few2012ramses}, the star formation recipe used in Illustris is an effective equation of state, where stars form above a certain gas density of ($\rho_{SFR}$) with a star formation time scale ($t_{SFR}$). The values used in Illustris are: $\rho_{SFR}=0.13$ cm$^{-3}$ and $t_{SFR} = 2.2$ Gyr. The stochastic prescription for star formation follows the Kennicutt-Schmidt law and adopts a Chabrier initial mass function. 

\par
The cooling rate of gas in Illustris is calculated as a function of gas density, temperature, metallicity, the radiation fields of AGN, and the ionising background radiation from galaxies and quasars. The primordial cooling is calculated and combined with the cooling due to metals, by using CLOUDY \citep{ferland1998cloudy}. When a stellar particle is formed in the Illustris simulation, it inherits the metallicity of the local gas. Star particles slowly return mass to the interstellar medium to account for mass loss from aging stellar populations. Stellar feedback is modelled through a wind scheme with a velocity scaling based on the local one-dimensional dark matter $V_{disp}$ $(3.7*V_{disp})$.
\par
Every FoF group above $M > 7.1 \times 10^7 M_{\odot}$ is seeded with a supermassive black hole with mass $1.4 \times 10^5 M_{\odot}$. Illustris includes treatments for both quasar mode and radio mode feedback \citep{sijacki2014illustris}, depending on what the accretion rate is at a particular time. The mass accretion is defined by a Bondi-Hoyle-Lyttleton based Eddington-limited rate given by:
\begin{equation}
\dot M_{BH} = min\left[\frac{4\pi\alpha G^2 M_{BH}^2 \rho}{(c_s^2 + v_{BH}^2)^{\frac{3}{2}}}, \dot M_{edd}\right]
\end{equation} 
A novel prescription for radiative AGN feedback is implemented by assuming an average AGN SED and a bolometric, luminosity-dependent scaling; this model is further described in \citet{vogelsberger2013model, sijacki2014illustris}. 

\par 
For this work, we extracted the SUBFIND catalogs \citep{nelson2015illustris} at four epochs: $z=0, 1, 2, 4$ from \href{http://illustris.org}{\texttt{http://illustris.org}}. The following dark matter (sub)halo properties were used as the inputs to our ML algorithms: $M_{DM}$, $S_{x}$, $S_{y}$, $S_{z}$, $V_{disp}$, $V_{circular}$, and $\mathcal{N}_{DM}$. $M_{DM}$ is the total mass of the subhalo, $S_{x,y,z}$ refers to the different components of the spin, $V_{disp}$ denotes the velocity dispersion, $V_{circular}$ is the maximum circular velocity in the subhalo, and $\mathcal{N}_{DM}$ is the number of DM particles bound to the subhalo. We use these to predict the following attributes that accumulate through cosmic time due to a wide range of processes: $M_{gas}$, $M_{\star}$, $M_{BH}$, $SFR$, metallicity, and $M_{\star, half}$. $M_{gas}$ is the total gas mass in the subhalo. To make our training and testing set, we pose two constrains: $M_{dm} > 10^9 M_{\odot} h^{-1}$ and $M_{star} \geq 0$. No minimum on the number of stellar particles was imposed to end up with a more dynamic range of masses; the larger range enables us to apply results from a lower resolution simulation to a higher resolution simulation. The number of (sub)haloes at $z=0,1,2,4$ are: $249370, 342622, 433406,$ and $403268$ respectively. 
The training sample, $25\%$ of the catalog, was randomly chosen from the Illustris simulation. The other $75\%$ was used for the testing and the comparisons shown in Section \ref{gp}.

\par 
In the growing climate of scientific reproducibility, we follow the trend of making all out data code available at: \href{http://github.com/ProfessorBrunner/ml-sims}{\texttt{http://github.com/ProfessorBrunner/ml-sims}}. 

\subsection{Machine Learning}
In this section, we briefly discuss the basics of machine learning and we also provide the pseudocode for and briefly discuss the ML algorithm that performed the best in our analyses. 
\subsubsection{General Overview}
Machine learning is a popular field in computer science, with
a wide variety of applications in a number of other areas.
The basic idea of ML algorithms is to `learn' approximate relationships between the input data and the output data without any explicit analytical prescription being used. Supervised learning techniques are provided some training data (X, y) and they try to learn the mapping G(X $\rightarrow$ y) in order to apply this mapping to the test data.

\par Machine learning has been applied to several subfields
in Astronomy with a lot of success; see, for example, \citet{ball2010data, ivezic2014statistics}. A majority of the applications of ML in astronomy have either been in classification problems such as star-galaxy classification
\citep{ball2006robust, kim2015hybrid}, galaxy morphology classification \citep{banerji2010galaxy, dieleman2015rotation} or have
been regression applications like: photometric redshift estimation \citep{ball2007robust, gerdes2010arborz, kind2013tpz}, estimation of stellar atmospheric parameters \citep{fiorentin2007estimation}, and determining stellar labels from spectroscopic data \citep{ness2015cannon}.

\par 
We use various statistics later in the paper to quantify the effectiveness of ML in predicting the galaxy properties. First, we use the standard mean squared error (MSE) metric, which is defined as: 
\begin{equation}
MSE = \frac{1}{N_{test}}\sum_{i=1}^{i=N_{test}-1} \left(X_{test}^{i} - X_{predicted}^{i}\right)^2
\end{equation} 
Here, $X_{test}^i$ is the $i^{th}$ value of the actual test set and $X_{predicted}^i$ is the $i^{th}$ value of the predicted set. Furthermore, to quantify the effectiveness of the ML algorithms, we use the base MSE ($MSE_b$) defined as:
\begin{equation}
MSE_b = \frac{1}{N_{test}}\sum_{i=1}^{i=N_{test}-1} \left(X_{test}^{i} - X_{mean,train}\right)^2
\end{equation}
Here, $X_{mean,train}$ is the mean of the training data set. $MSE_b$ is an extremely naive prediction of the error since each test point is simply predicted as the mean of the training dataset. The ratio of  $MSE_b$ to $MSE$ is an indicator of the goodness of fit. The value of this fraction very much depends on the distribution of the quantity under consideration. If the mean is a good predictor for the quantity being considered, $MSE_b$ will be very low as well leading to a smaller ratio.Our motivation in using this metric was supplementary; the primary indicators of goodness of fit are $R^2$ and Pearson correlation, which are defined below.

\par 
We will also be using the following two metrics to check for the robustness of the prediction: the Pearson correlation and the coefficient of determination (`regression score'). The Pearson correlation is written as:
\begin{equation}
\rho = \frac{cov(X_{predicted}, X_{test})}{\sigma_{X_{predicted}} \sigma_{X_{test}}}
\end{equation}
 and $R^2$ as: 
 \begin{equation}
 R^2 = 1 - \frac{\sum_i (X_{test}^i - X_{predicted}^i)^2}{\sum_i (X_{test}^i - X_{mean,train})^2}
 \end{equation}
 
\begin{table*}
\begin{minipage}{140mm}
 \caption{An outline of the extremely randomized trees regression algorithm}
 \label{symbols}
 \begin{tabular}{@{}lcccccc}
  \hline
  \hline
  \begin{centering}
  Extremely Randomized Trees
  \end{centering} \\
  \hline
  \textbf{Inputs}: A training set $S$ corresponding to \textbf{$(X, y)$} input-output vectors, where \\ \textbf{X}=$(X_1,X_2,...,X_N)$ and  \textbf{y}=$(y_1,y_2,...,y_l)$, M (number of trees in the ensemble), \\ K (number of random splits screened at each node) and $n_{min,samples}$ (number of samples \\ required to split a node)\\
  \textbf{Outputs}: An ensemble of M trees: $\mathcal{T} = (t_1, t_2,...,t_M)$ \\
  \hline
  \textit{\textbf{Step 1}}: Randomly select $K$ inputs $(X_1,X_2,...,X_K)$ where $1\leq K \leq N)$. \\
  \hline 
  \textit{\textbf{Step 2}}: For each selected input variable $X_i$ in $i=(1,2,...,K)$: \\
  \tabitem Compute the minimal and maximal value of $X$ in the set: $X_i^{min}$ \\ and $X_i^{max}$ \\
  \tabitem Randomly select a cut-point $X_c$ in the interval [$X_i^{min}$, $X_i^{max}$] \\
  \tabitem Return the split in the interval $X_i \leq X_c$ \\
  \hline
  \textit{\textbf{Step 3}}: Select the best split $s_{\star}$ such that Score($s_{\star}$, S) = $max_{i=1,2,...,K}$ Score($s_{i}$, S) \\
  \hline
  \textbf{\textit{Step 4}}: Using $s_{\star}$, split $S$ into $S^l(X_i)$  and $S^r(X_i)$ \\
  \hline
  \textbf{\textit{Step 5}}: For $S^l(X_i)$  and $S^r(X_i)$, check the following conditions: \\
  \tabitem $|S^l(X_i)|$ or $|S^r(X_i)|$ is lower than $n_{min,samples}$ \\
  \tabitem All input attributes $(X_1,X_2,...,X_N)$ are constant in $|S^l(X_i)|$ or $|S^r(X_i)|$ \\
  \tabitem The output vector $(y_1,y_2,...,y_l)$ is constant in $|S^l(X_i)|$ or $|S^r(X_i)|$ \\
  \hline
  \textbf{\textit{Step 6}}: If any of the conditions in step 5 are satisfied, stop. We're at a leaf node. \\
  If none of the conditions are satisfied, repeat steps 1 through 5. \\
  \hline 
\end{tabular}
\end{minipage}
\end{table*}

\subsubsection{Extremely Randomized Trees} 
The primary algorithm used in this work was extremely randomized trees \citep{geurts2006extremely} (ERT). ERT is an ensemble technique that build on a weak learner (decision trees, in this case). 

\par
The essence of ERT is to build a large ensemble of regression trees where both the attribute and split-point choice are randomized while splitting a tree node. We provide pseudocode for the full algorithm in Table \ref{symbols}, which closely follows the algorithm outlined in \citet{geurts2006extremely}. In the algorithm, the Score is the reduction in the variance. For the two subtrees $S^l$  and $S^r$ corresponding to the split $s_{\star}$, the Score($s_{\star}$, S), abbreviated to $Sc(s_{\star}, S)$, is given by:
\begin{equation}
Sc(s_{\star}, S) = \frac{var(\textbf{y}, S) - \frac{|S^l|}{|S|} var(\textbf{y}, S^l) - \frac{|S^r|}{|S|} var(\textbf{y}, S^r)}{var(\textbf{y},S)}
\end{equation}

The estimates produced by the $M$ trees in the ERT ensemble are finally combined by averaging $y$ over all trees in the ensemble. The use of the original training data set in place of a bootstrap sample (as is done for random forests) is done to minimize bias in the prediction. Furthermore, the use of both randomization and averaging is aimed at reducing the variance of the prediction \citep{geurts2006extremely}. The added randomness corresponds to uncorrelated errors and mitigates bias in the predictions. 
We used the implementation of ERT found in the Python package, scikit-learn \citep{pedregosa2011scikit}.

\section{Results} \label{rd}
In this section, we present our results of applying ML techniques to the Illustris simulation. The following attributes are predicted at multiple epochs: gas mass, stellar mass, black hole mass, stellar mass inside the half-mass radius, star formation rate, metallicity, and $g-r$ color. We perform our analysis for $z=0, 1, 2, 4$ and present our results in this paper for $z=0$ and $z=2$. The rest can be found in the linked Github repository. We also exclude the stellar half mass results as these are incredibly similar to the stellar mass results. 

\subsection{Galaxy Properties} \label{gp}
\par
The structure of this section is as follows. For each physical attribute that is predicted, we provide a table summarizing some basic statistical quantities that are indicative of the goodness of fit at $z=0, 2$. $MSE_b$ and the $MSE$ are listed for each technique. The factor reduction of the $MSE$ is also listed to test the relative performance of the algorithms to quantify how much they are learning. Finally, the Pearson correlation between the predicted and the true data set and the coefficient of determination ($R^2$) are also listed. We provide a hexbin plot (on a log scale) of the predicted quantity versus the quantity in Illustris. For the hexbin plot, a gridsize of 30 was used and the colormap was logarithmically scaled. A violinplot is also shown to compare the distributions of a particular physical attribute in Illustris galaxies and the ML galaxies. To provide an object-by-object comparison, for both redshifts, a plot showing the number density of galaxies as a function of the attribute is shown for both the Illustris and ML simulated galaxies and a plot showing the ratio of $attribute_{ML}$ to $attribute_{Illustris}$ is shown as a function of $attribute_{Illustris}$. We also provide, when appropriate, a plot showing how the ML and Illustris galaxies follow a given observational constraint. 

\subsubsection{Gas Mass}
\begin{figure*}
  \includegraphics[width=168mm]{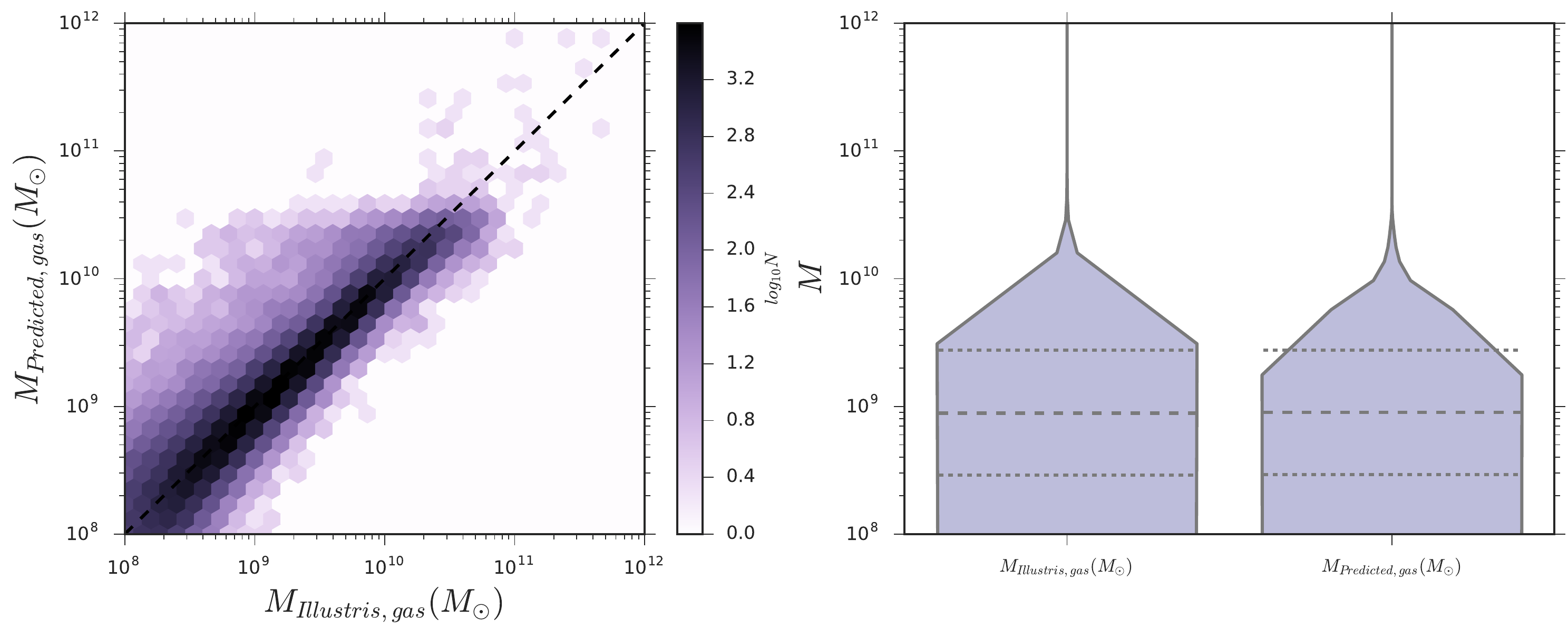}
    \caption{\textbf{\textit{Left}}: A hexbin plot of $M_{Illustris,gas}$ and $M_{predicted,gas}$ at $z=0$. The black dashed line corresponds to a perfect prediction. \textbf{\textit{Right}}: A violinplot showing the distributions of $M_{Illustris,gas}$ and $M_{predicted,gas}$. The median and the interquantile range are also shown.}
        \label{figure1}
\end{figure*}

\begin{figure*}
  \includegraphics[width=168mm]{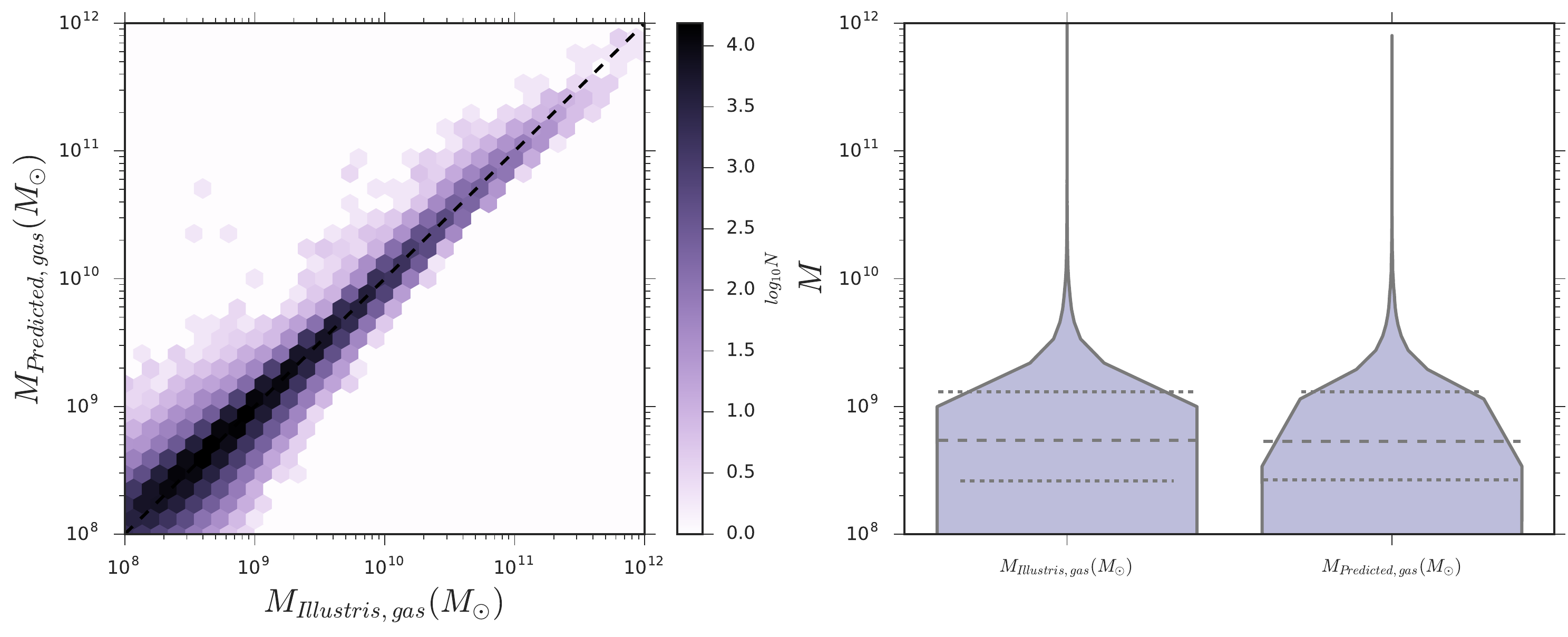}
    \caption{\textbf{\textit{Left}}: A hexbin plot of $M_{Illustris,gas}$ and $M_{predicted,gas}$ at $z=2$. The black dashed line corresponds to a perfect prediction. \textbf{\textit{Right}}: A violinplot showing the distributions of $M_{Illustris,gas}$ and $M_{predicted,gas}$. The median and the interquantile range are also shown.}
   \label{f2}
\end{figure*}

\begin{figure*}
  \includegraphics[width=168mm]{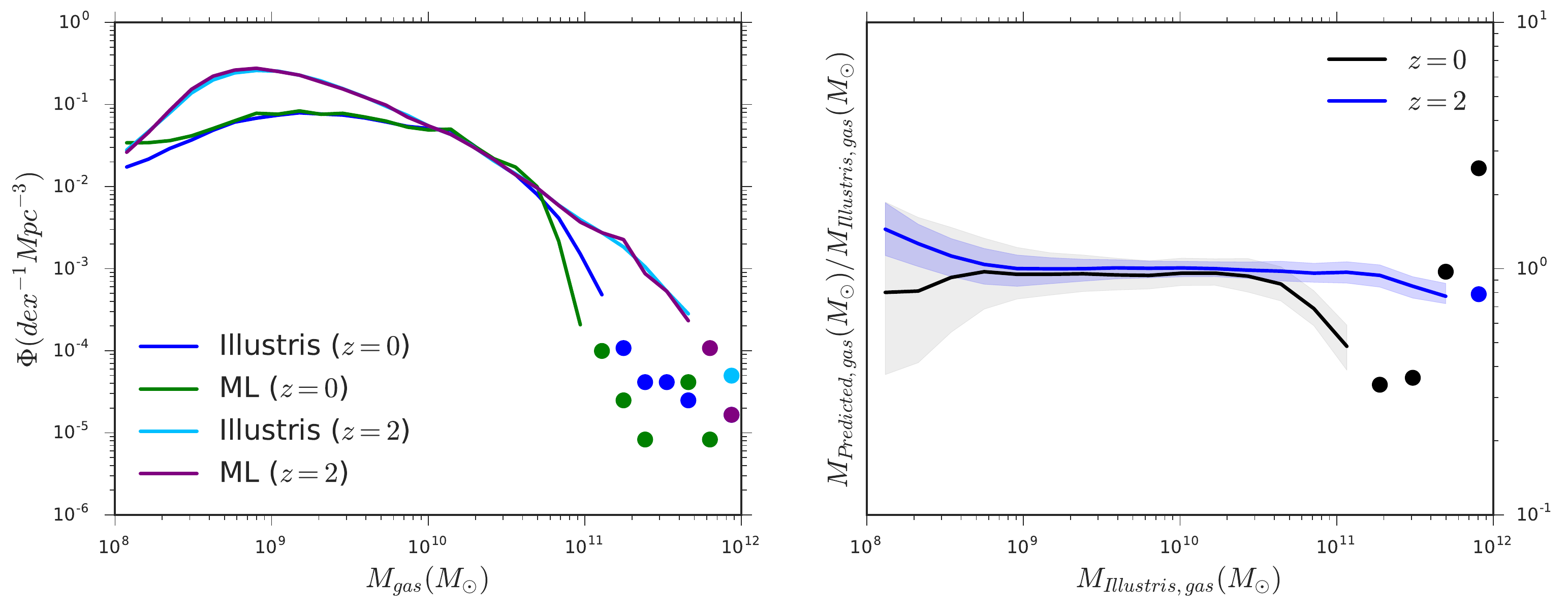}
    \caption{\textbf{\textit{Left}}: The gas mass function for the Illustris and the ML simulated galaxies at both $z=0$ and $z=2$. Bins with less than 20 galaxies are denoted with big dots. \textbf{\textit{Right}}: A binned plot of the fraction $\frac{M_{Illustris,gas}}{M_{predicted,gas}}$ as a function of $M_{Illustris,gas}$. The area between the 25th and the 75th percentile for each bin has been shaded. The curve is dashed when there are less than 20 galaxies in the mass bin.}
   \label{f23}
\end{figure*}

In Table \ref{table2} and Figures \ref{figure1}, \ref{f2}, and \ref{f23}, the results for the gas mass prediction are shown. There are several physical processes that play a role in how gas mass is accumulated through cosmic time. Changes in the gas mass are driven by four key processes: gas cooling, cosmic accretion, feedback, and turning into stars and stars returning gas mass through gas recycling. As we can see in the two plots and the table provided, ML is able to approximate the combined effects of these processes reasonably well. 

\par In Figure \ref{f23}, the gas mass function for the Illustris and the ML simulated galaxies at both $z=0$ and $z=2$ is shown alongside a binned plot of the fraction $\frac{M_{predicted,gas}}{M_{Illustris,gas}}$ as a function of $M_{Illustris,gas}$. The mass function plot shows that the ML prediction at both epochs matches up very well with the Illustris galaxies. Furthermore, the fraction plot serves as a good object-by-object comparison of the ML and Illustris galaxies. The binned fraction plot shown in Figure \ref{f23} shows high scatter at lower masses, which is consistent with what is shown in the hexbin plot, but shows an average value of $1$ for the ratio of $M_{predicted,gas}$ to $M_{Illustris,gas}$. The statistically consistent results (from the hexbin plots) along with the mass function and the fraction plots imply that the population of ML simulated galaxies is consistent with the population of Illustris galaxies at both epochs. 

\begin{table}
\caption{Gas mass statistics}
 \label{table2}
  \begin{tabular}{@{}lcccccc}
   Redshift & $MSE_b$ & $MSE$ & ($\frac{MSE_b}{MSE}$) & $\rho$
         & $R^2$\\
  \hline
   $z=0$ & 20.294 & 6.641 & 3.055 & 0.849 & 0.673\\
   \hline
   $z=2$ & 0.761 & 0.063 & 12.002 & 0.959 & 0.917\\
  \hline
 \end{tabular}
\end{table}

\subsubsection{Stellar Mass}

\begin{figure*}
  \includegraphics[width=168mm]{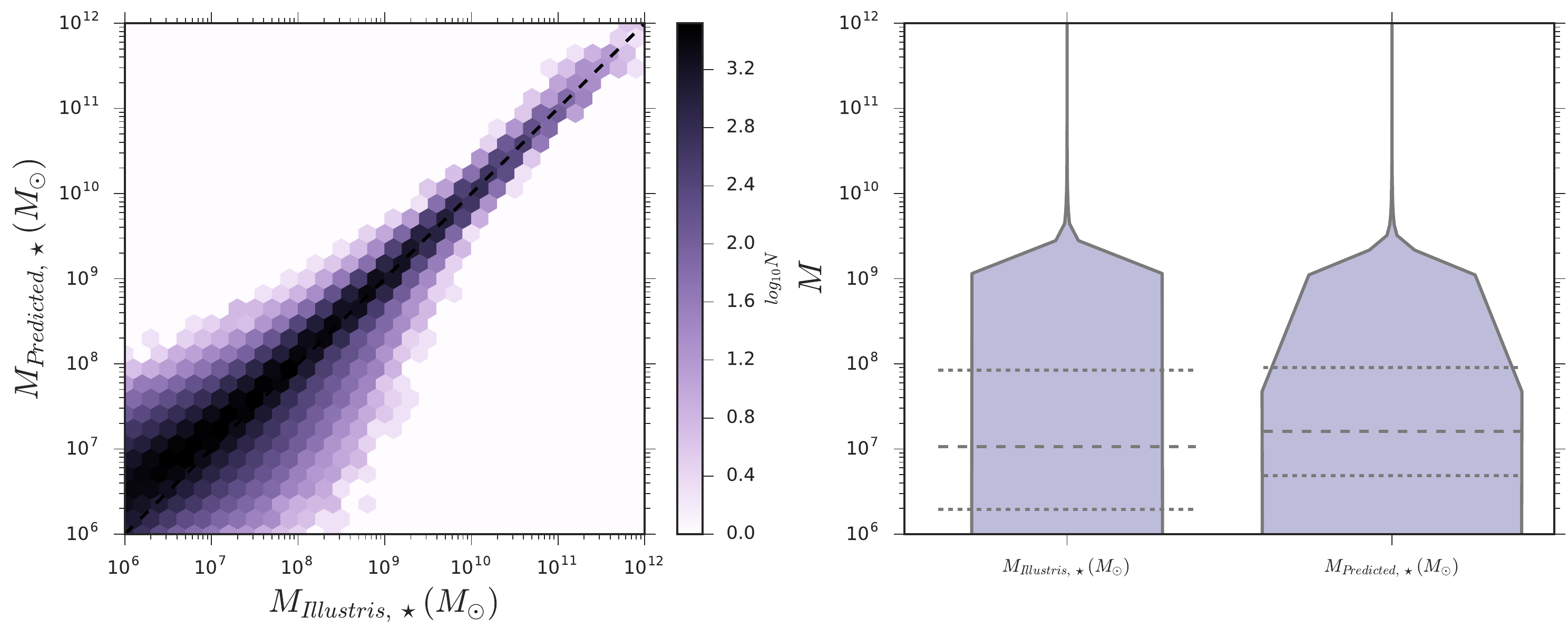}
    \caption{\textbf{\textit{Left}}: A hexbin plot of $M_{Illustris,\star}$ and $M_{predicted,\star}$ at $z=0$. The black dashed line corresponds to a perfect prediction. \textbf{\textit{Right}}: A violinplot showing the distributions of $M_{Illustris,\star}$ and $M_{predicted,\star}$. The median and the interquantile range are also shown.}
        \label{f3}
\end{figure*}

\begin{figure*}
  \includegraphics[width=168mm]{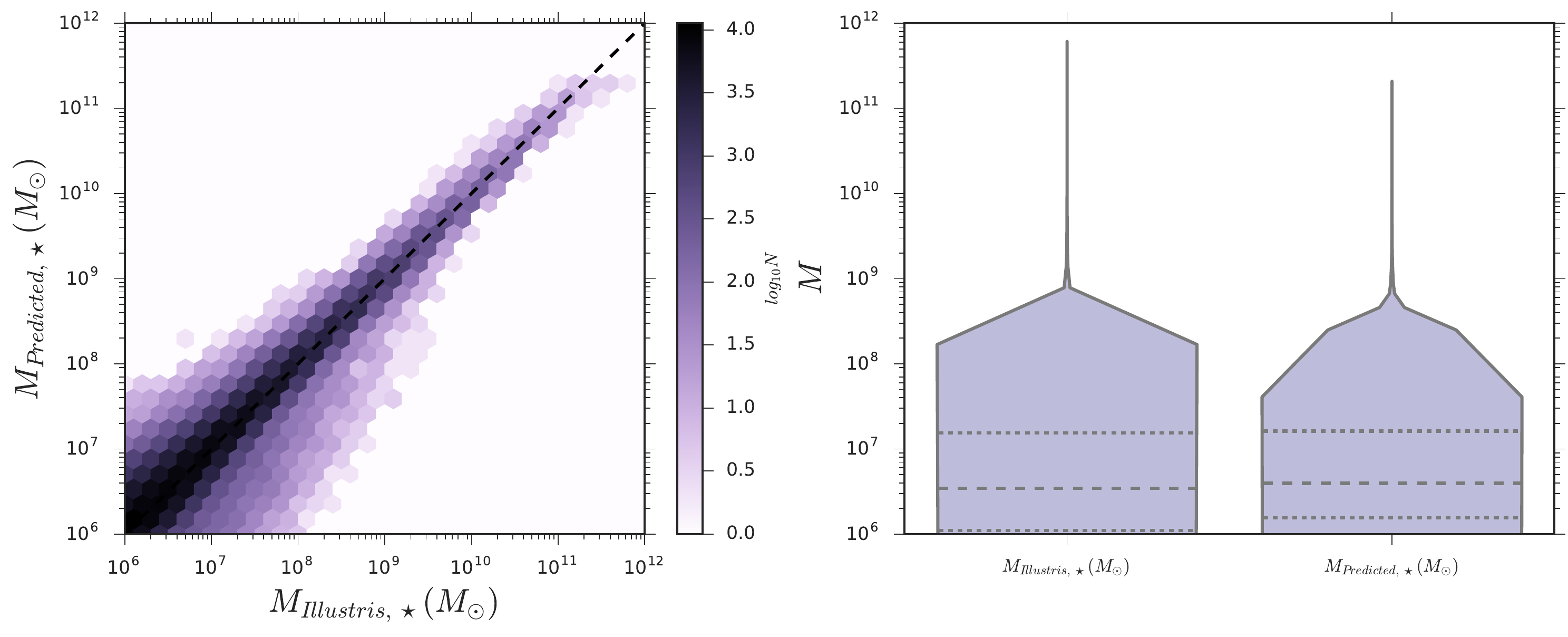}
    \caption{\textbf{\textit{Left}}: A hexbin plot of $M_{Illustris,\star}$ and $M_{predicted,\star}$ at $z=2$. The black dashed line corresponds to a perfect prediction. \textbf{\textit{Right}}: A violinplot showing the distributions of $M_{Illustris,\star}$ and $M_{predicted,\star}$. The median and the interquantile range are also shown.}    
    \label{f4}
\end{figure*}

\begin{figure*}
  \includegraphics[width=168mm]{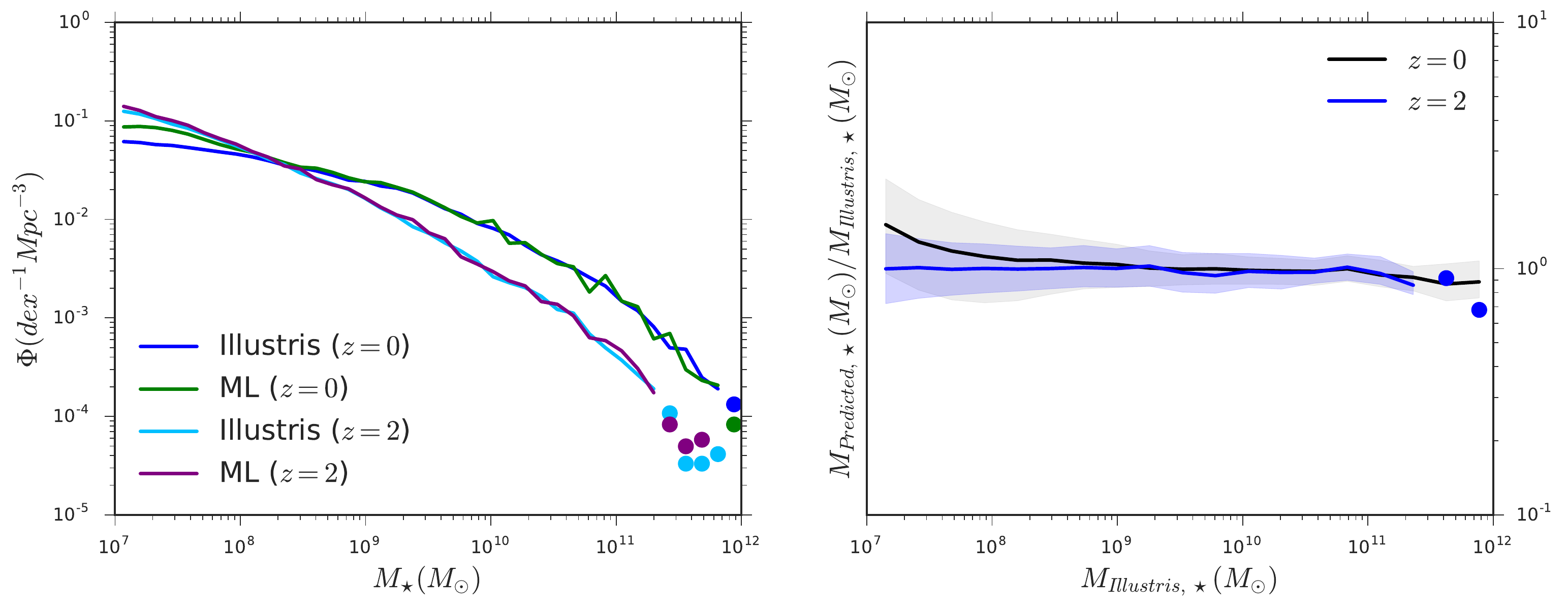}
    \caption{\textbf{\textit{Left}}: The stellar mass function for the Illustris and the ML simulated galaxies at both $z=0$ and $z=2$. \textbf{\textit{Right}}: A binned plot of the fraction $\frac{M_{Illustris,\star}}{M_{predicted,\star}}$ as a function of $M_{Illustris,\star}$. The area between the 25th and the 75th percentile for each bin has been shaded. Bins with less than 20 galaxies are denoted with big dots.}   
    \label{f26}
\end{figure*}

The results for the stellar mass prediction are shown in Table \ref{t3} and Figures \ref{f3}, \ref{f4}, \ref{f26}, \ref{f18}, and \ref{f19}. The stellar mass prediction, like the gas mass, is being predicted very well. The buildup of stellar mass in Illustris occurs when a gas cell exceeds some critical density $\rho_{sfr}$; when this condition is met, star particles are produced with a timescale $t_{sfr}$ using free parameters that are fine-tuned to follow the Kennicutt-Schmidt relation. Our results for the stellar mass using solely halo properties are promising and imply that ML is able to model the combined effects of the recipes used to accumulate stellar mass very well.

\par 
However, there is still some scatter in the predicted results. In the case of the stellar mass, the AGN feedback and the supernovae feedback quench star formation (especially for higher mass haloes) and a purely dark matter based phenomenological is unable to model these phenomena. The high $R^2$ values and the hexbin plot place confidence in our predictions. Moreover, the distribution of the stellar mass is reproduced almost perfectly as shown in Figure 4. The results shown are statistically robust and constitute a set of galaxies with stellar masses that is similar to that found in Illustris. 

\par In Figure \ref{f26}, the stellar mass function for the Illustris and the ML simulated galaxies at both $z=0$ and $z=2$ is shown alongside a binned plot of the fraction $\frac{M_{predicted,\star}}{M_{Illustris,\star}}$ as a function of $M_{Illustris,\star}$. The mass function plot shows that the ML prediction at both epochs matches up very well with the Illustris galaxies. Furthermore, the fraction plot serves as a good object-by-object comparison of the ML and Illustris galaxies. The fraction plot shown in Figure \ref{f26} shows high scatter at lower masses, which is consistent with what is shown in the hexbin plot, but shows an average value of $1$ for the ratio of $M_{predicted,\star}$ to $M_{Illustris,\star}$. The statistically consistent results (from the hexbin plots) along with the mass function and the fraction plots imply that the population of ML simulated galaxies is consistent with the population of Illustris galaxies at both epochs.

\par
Furthermore, in Figures \ref{f18} and \ref{f19}, we show the stellar mass-halo mass relation at $z=0$ and $z=2$. The SMHM is reproduced almost perfectly at lower masses. There are some deviations at higher masses, but we posit that this is because the number of dark matter haloes with such a high mass is low, thereby leading to a smaller sample size and some deviations from expectations.  Therefore, the discrepancy at higher masses can be attributed to the lack of sufficient training data for the ML algorithms. The SMHM being reproduced is incredibly promising because it shows that ML is able to approximate the mapping between dark matter haloes and stellar mass buildup very well. Another point to note here, that we will come back to later in the paper, is that there's no direct relationship being input or assumed by ML like SHAMs; instead, a relationship is approximated from the results of an NBHS to predict how dark matter haloes are populated with galaxies. 

\begin{table}
\caption{Stellar Mass statistics}
 \label{t3}
 \begin{tabular}{@{}lcccccc}
   Redshift & $MSE_b$ & $MSE$ & ($\frac{MSE_b}{MSE}$) & $\rho$
         & $R^2$\\
  \hline
   $z=0$ & 1.506 & 0.126 & 11.928 & 0.957 & 0.916\\
   \hline
   $z=2$ & 0.081 & 0.011 & 7.486 & 0.936 & 0.866\\
  \hline
 \end{tabular}
\end{table}

\subsubsection{Black Hole Mass}
\begin{figure*}
  \includegraphics[width=168mm]{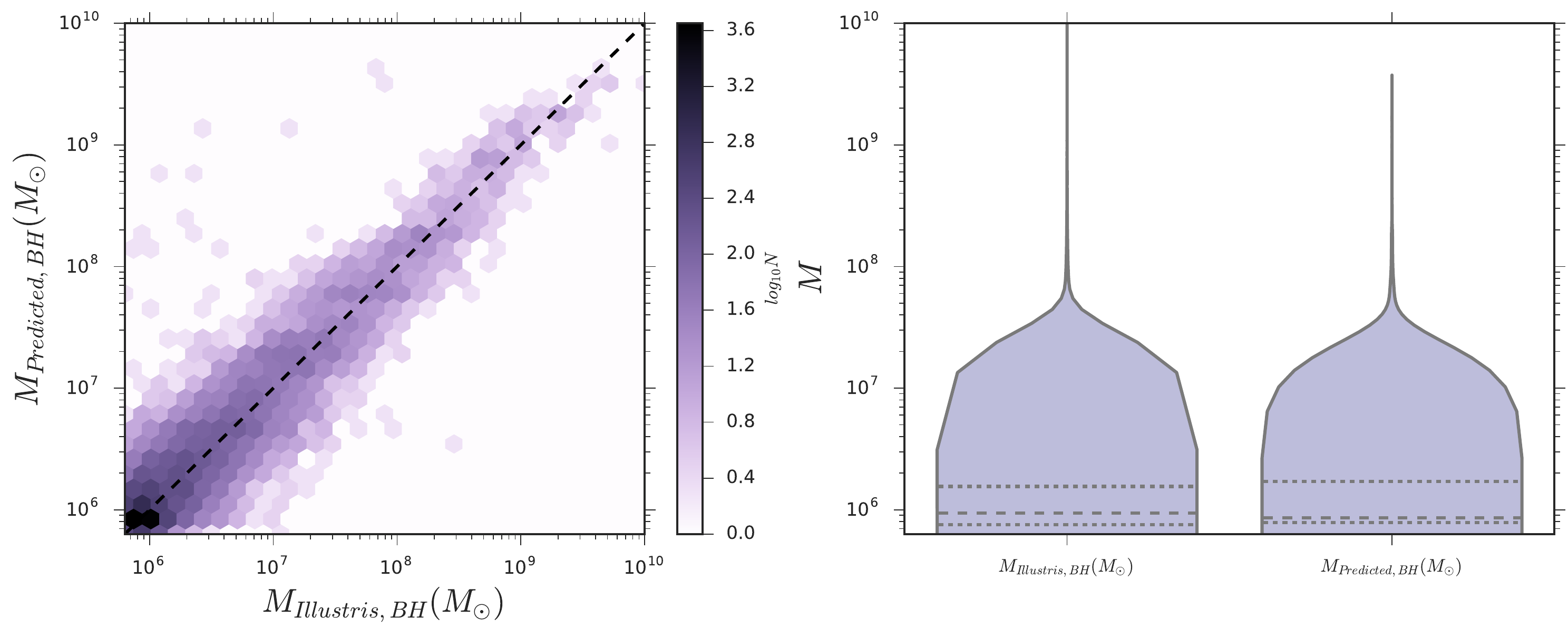}
    \caption{\textbf{\textit{Left}}: A hexbin plot of $M_{Illustris,BH}$ and $M_{predicted,BH}$ at $z=0$. The black dashed line corresponds to a perfect prediction. \textbf{\textit{Right}}: A violinplot showing the distributions of $M_{Illustris,BH}$ and $M_{predicted,BH}$. The median and the interquantile range are also shown.}
    \label{f5}
\end{figure*}

\begin{figure*}
  \includegraphics[width=168mm]{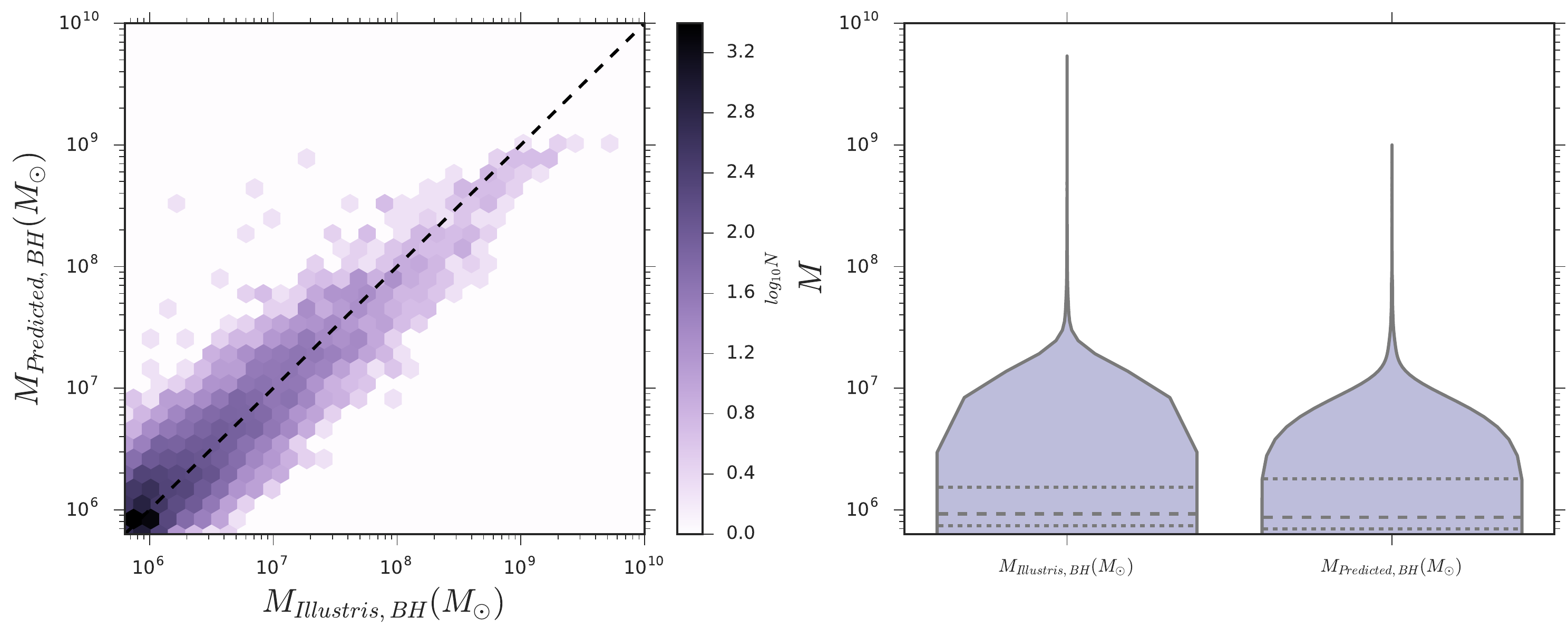}
        \caption{\textbf{\textit{Left}}: A hexbin plot of $M_{Illustris,BH}$ and $M_{predicted,BH}$ at $z=2$. The black dashed line corresponds to a perfect prediction. \textbf{\textit{Right}}: A violinplot showing the distributions of $M_{Illustris,BH}$ and $M_{predicted,BH}$. The median and the interquantile range are also shown.}
    \label{f6}
\end{figure*}

\begin{figure*}
  \includegraphics[width=168mm]{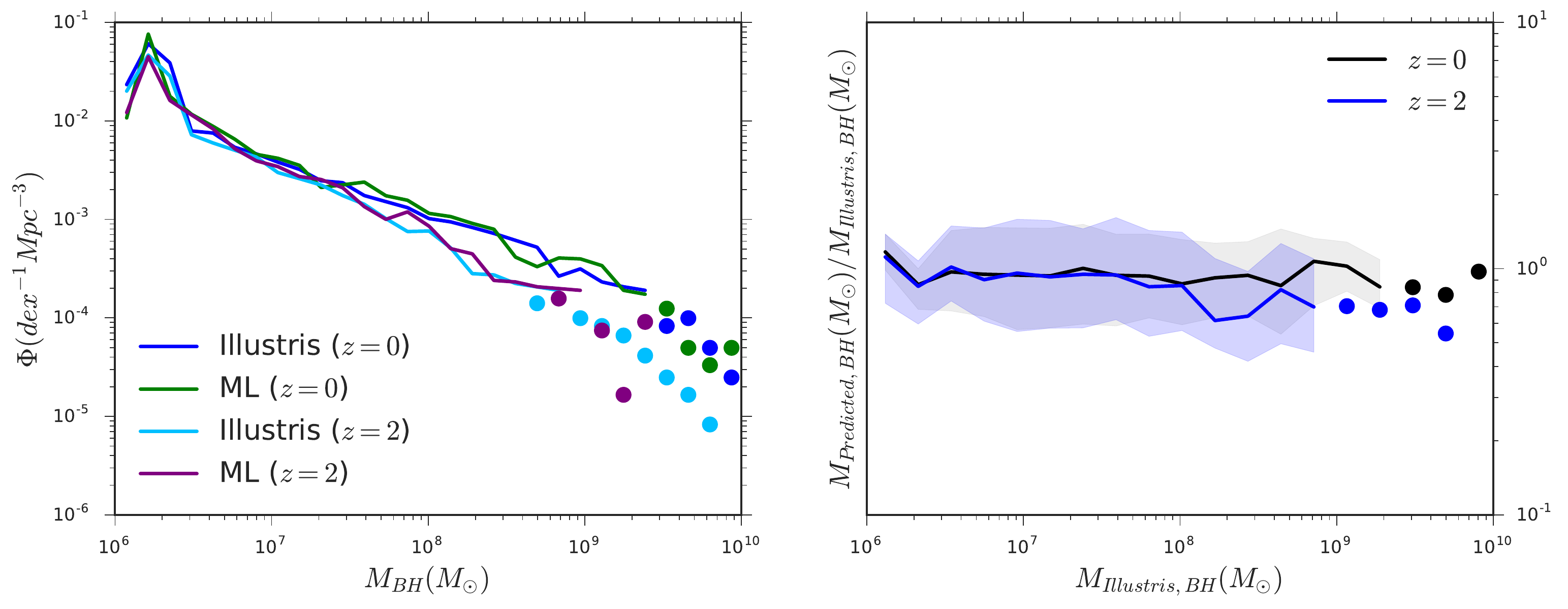}
  \caption{\textbf{\textit{Left}}: The BH mass function for the Illustris and the ML simulated galaxies at both $z=0$ and $z=2$. \textbf{\textit{Right}}: A binned plot of the fraction $\frac{M_{Illustris,BH}}{M_{predicted,BH}}$ as a function of $M_{Illustris,BH}$. The area between the 25th and the 75th percentile for each bin has been shaded. Bins with less than 20 galaxies are denoted with big dots.}
    \label{f21}
\end{figure*}

The results for the black hole mass, which is also reproduced very well, are shown in Table \ref{t4} and Figures \ref{f5}, \ref{f6}, \ref{f21}, \ref{f12}, and \ref{f13}. In Section \ref{db}, we discussed how the black hole mass is accreted in Illustris. As with other quantities, the hexbin plot shows some scatter for lower masses but as the mass increases, the predictions and simulated values match up very well. Our results for the black hole mass using solely halo properties are promising and imply that ML is able to model combined effects of the recipes used to accumulate black hole mass in Illustris. 

\par In Figure \ref{f21}, the BH mass function for the Illustris and the ML simulated galaxies at both $z=0$ and $z=2$ is shown alongside a binned plot of the fraction $\frac{M_{predicted,BH}}{M_{Illustris,BH}}$ as a function of $M_{Illustris,\star}$. The mass function plot shows that the ML prediction at both epochs matches up well with the Illustris galaxies. Furthermore, the fraction plot serves as a good object-by-object comparison of the ML and Illustris galaxies. The fraction plot shown in Figure \ref{f21} shows high scatter at lower masses, which is consistent with what is shown in the hexbin plot, but shows an average value of $1$ for the ratio of $M_{predicted,BH}$ to $M_{Illustris,BH}$. The statistically consistent results (from the hexbin plots) along with the mass function and the fraction plots imply that the population of ML simulated galaxies is consistent with the population of Illustris galaxies at both epochs.

\par Furthermore, in Figures \ref{f12}, and \ref{f13}, the black hole mass-bulge mass relation is plotted. Following \citet{sijacki2014illustris}, we use $M_{\star,half}$ as proxy for the bulge mass. The prediction and the Illustris simulated relationship match up almost perfectly, serving as yet another indicator that ML is able to model the combined effects of the processes that play a role in BH mass accumulation reasonably well. The reproduction of the BH-bulge mass relationship further places confidence in the utility of ML to produce a global population of galaxies that is statistically robust. We can see in the figures that here is more scatter in the Illustris galaxies than in the ML galaxies. The reason for the scatter is discussed in detail in Section \ref{disc}. 

\begin{table}
\caption{Black Hole Mass statistics}
 \label{t4}
 \begin{tabular}{@{}lcccccc}
   Redshift & $MSE_b$ & $MSE$ & ($\frac{MSE_b}{MSE}$) & $\rho$
         & $R^2$\\
  \hline
   $z=0$ & $2.68 \times 10^{-5}$ & $7 \times 10^{-6}$ & 3.624  & 0.852 & 0.724\\
   \hline
   $z=2$ & $2.63 \times 10^{-6}$ & $1 \times 10^{-6}$ & 2.471 & 0.813 & 0.595\\
  \hline
 \end{tabular}
\end{table}

\subsubsection{Star Formation Rate}
\par
\begin{figure*}
  \includegraphics[width=168mm]{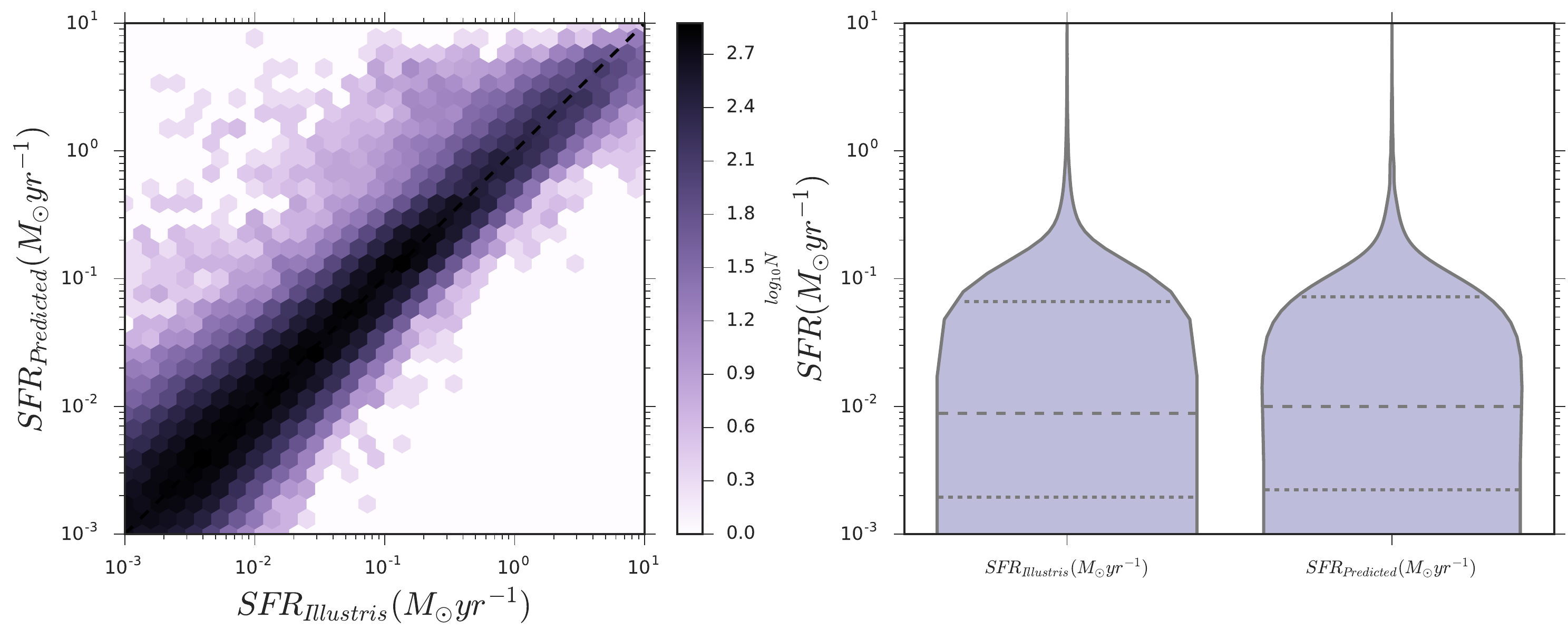}
        \caption{\textbf{\textit{Left}}: A hexbin plot of $SFR_{Illustris}$ and $SFR_{predicted}$ at $z=0$. The black dashed line corresponds to a perfect prediction. \textbf{\textit{Right}}: A violinplot showing the distributions of $SFR_{Illustris}$ and $SFR_{predicted}$. The median and the interquantile range are also shown.}
            \label{f7}

\end{figure*}

\begin{figure*}
  \includegraphics[width=168mm]{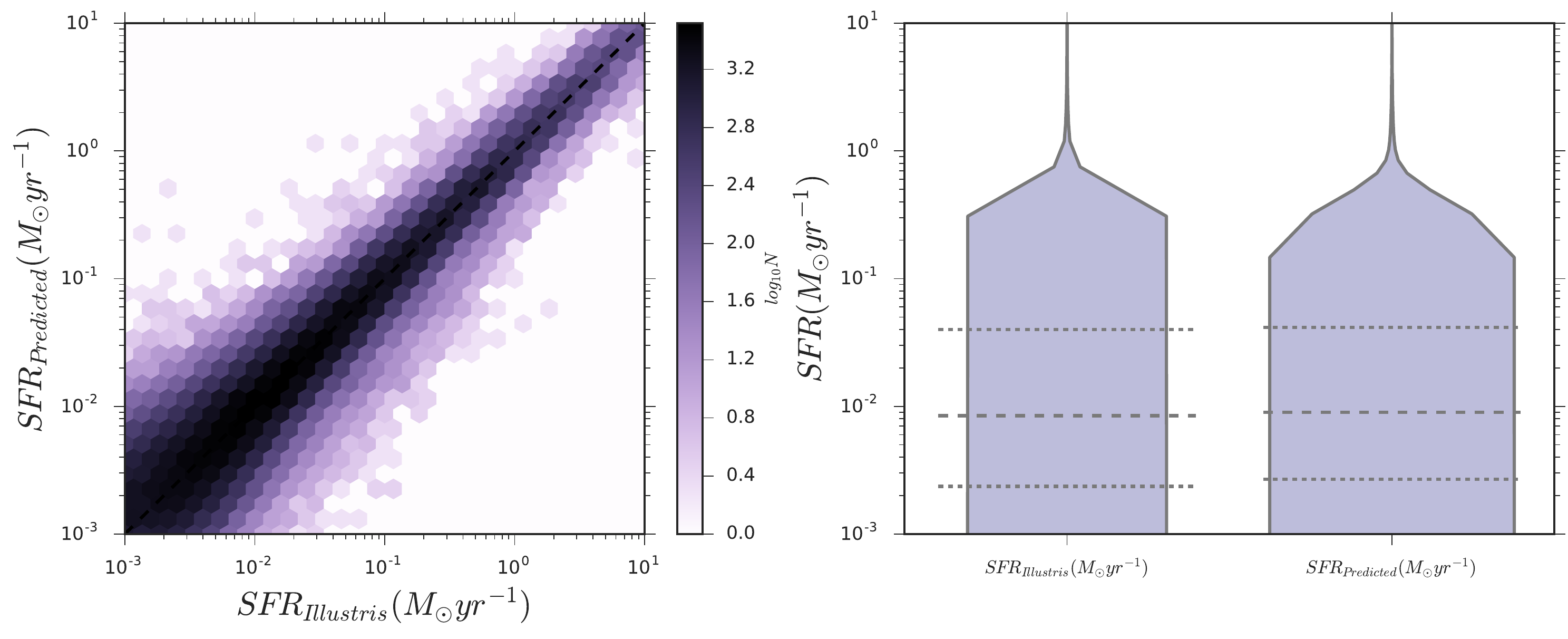}
    \caption{\textbf{\textit{Left}}: A hexbin plot of $SFR_{Illustris}$ and $SFR_{predicted}$ at $z=2$. The black dashed line corresponds to a perfect prediction. \textbf{\textit{Right}}: A violinplot showing the distributions of $SFR_{Illustris}$ and $SFR_{predicted}$. The median and the interquantile range are also shown.}
        \label{f8}
\end{figure*}

\begin{figure*}
  \includegraphics[width=168mm]{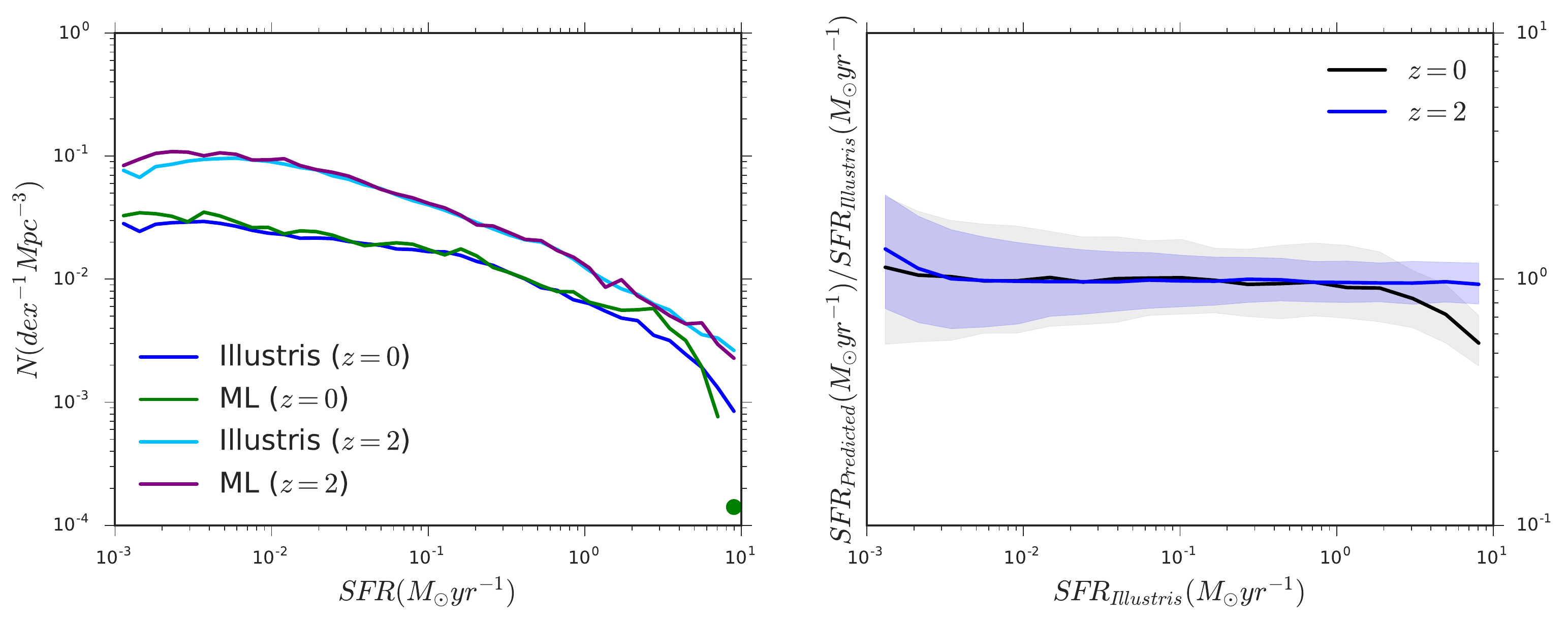}
\caption{\textbf{\textit{Left}}: The number density of galaxies as a function of SFR for the Illustris and the ML simulated galaxies at both $z=0$ and $z=2$.  \textbf{\textit{Right}}: A binned plot of the fraction $\frac{SFR_{Illustris}}{SFR_{predicted}}$ as a function of $SFR_{Illustris}$. The area between the 25th and the 75th percentile for each bin has been shaded. Bins with less than 20 galaxies are denoted with big dots.}
        \label{f25}
\end{figure*}

 In Table \ref{t5} and Figures \ref{f7}, \ref{f8}, \ref{f25}, \ref{f14}, \ref{f15}, \ref{f16}, and \ref{f17}, the results for the performance of ML at modeling the star formation rate using DM halo properties are presented. The statistics reported in Table 5 and the hexbin plot show that the predictions of the SFR are good but there is significantly more scatter compared to the other attributes. There is a noticeable overprediction, as evidenced by the hexbin plot and the violinplot. The distributions of the ML galaxies closely resembles that of the Illustris galaxies. At $z=2$, the predictions have significantly less scatter and the distributions match better. 

\par In Figure \ref{f25}, the number density of galaxies is plotted as a function of the SFR for the Illustris and the ML simulated galaxies at both $z=0$ and $z=2$ is shown alongside a binned plot of the fraction $\frac{SFR_{predicted}}{SFR_{Illustris}}$ as a function of $SFR_{Illustris}$. The number density plot shows that the ML prediction at both epochs matches up very well with the Illustris galaxies. Furthermore, the fraction plot serves as a good object-by-object comparison of the ML and Illustris galaxies. The fraction plot shown in Figure \ref{f25} shows higher scatter at lower values, which is consistent with what is shown in the hexbin plot, but shows an average value of $1$ for the ratio of the predicted $SFR$ to the Illustris $SFR$. The statistically consistent results (from the hexbin plots) along with the mass function and the fraction plots imply that the population of ML simulated galaxies is consistent with the population of Illustris galaxies at both epochs.

\par
Furthermore, in Figures \ref{f14} and \ref{f15}, the SFR is plotted as a function of stellar mass; the results for Illustris and the predicted results match up quite well at both epochs. The robust results imply that the set of galaxies, while it may not be numerically identical to Illustris, obeys a fundamental observational constraint. As with the earlier results, the standard deviation for the Illustris results is higher than the SFR predicted by ML. In Figures \ref{f16} and \ref{f17}, the specific SFR is plotted as a function of stellar mass. The results align very well at both epochs and demonstrate that ML is able to form an approximately good mapping between dark matter halo properties and the SFR.  

\begin{table}
\caption{SFR statistics}
 \label{t5}
 \begin{tabular}{@{}lcccccc}
   Redshift & $MSE_b$ & $MSE$ & ($\frac{MSE_b}{MSE}$) & $\rho$
         & $R^2$\\
  \hline
   $z=0$ & 0.377 & 0.140 & 2.702 & 0.794 & 0.630\\
   \hline
   $z=2$ & 10.565 & 2.754 & 3.836 & 0.865 & 0.739\\
  \hline
 \end{tabular}
\end{table}

\subsubsection{Stellar Metallicity}

\begin{figure*}
  \includegraphics[width=168mm]{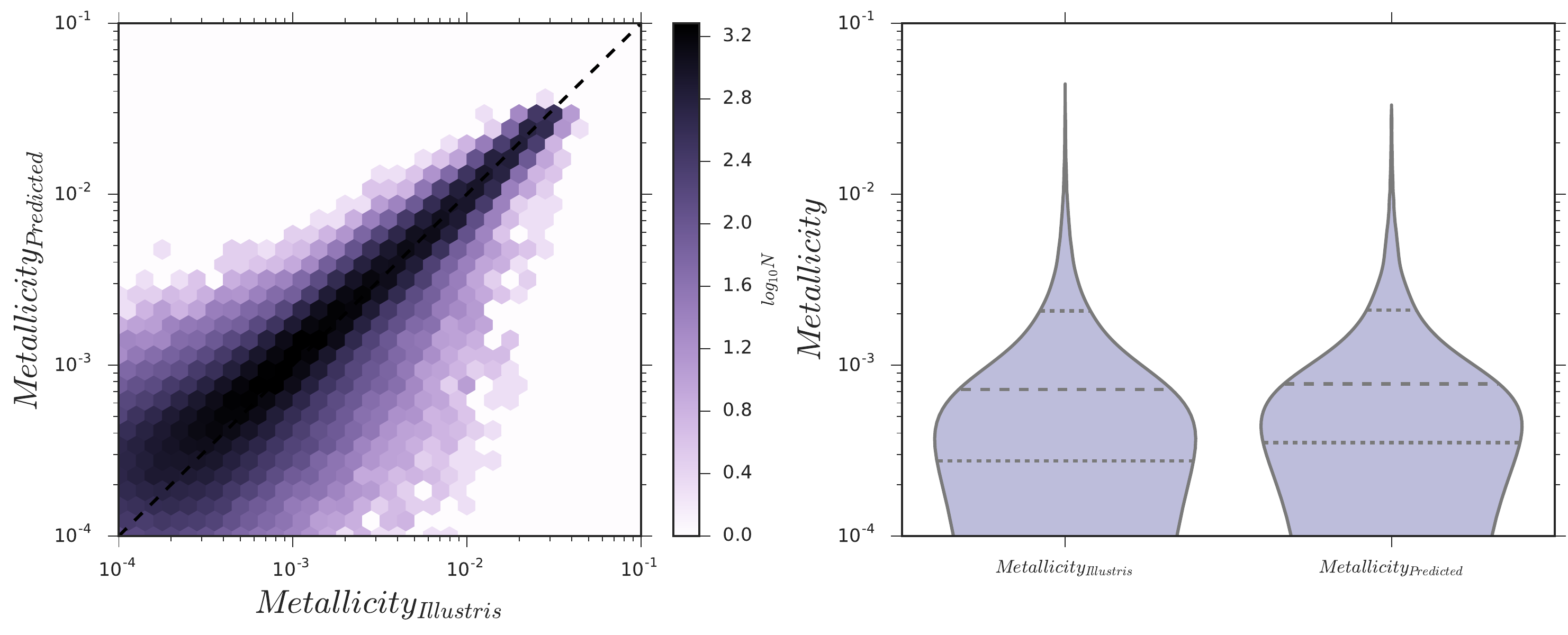}
    \caption{\textbf{\textit{Left}}: A hexbin plot of $Metallicity_{Illustris}$ and $Metallicity_{predicted}$ at $z=0$. The black dashed line corresponds to a perfect prediction. \textbf{\textit{Right}}: A violinplot showing the distributions of $Metallicity_{Illustris}$ and $Metallicity_{predicted}$. The median and the interquantile range are also shown.}
        \label{f9}

\end{figure*}

\begin{figure*}
  \includegraphics[width=168mm]{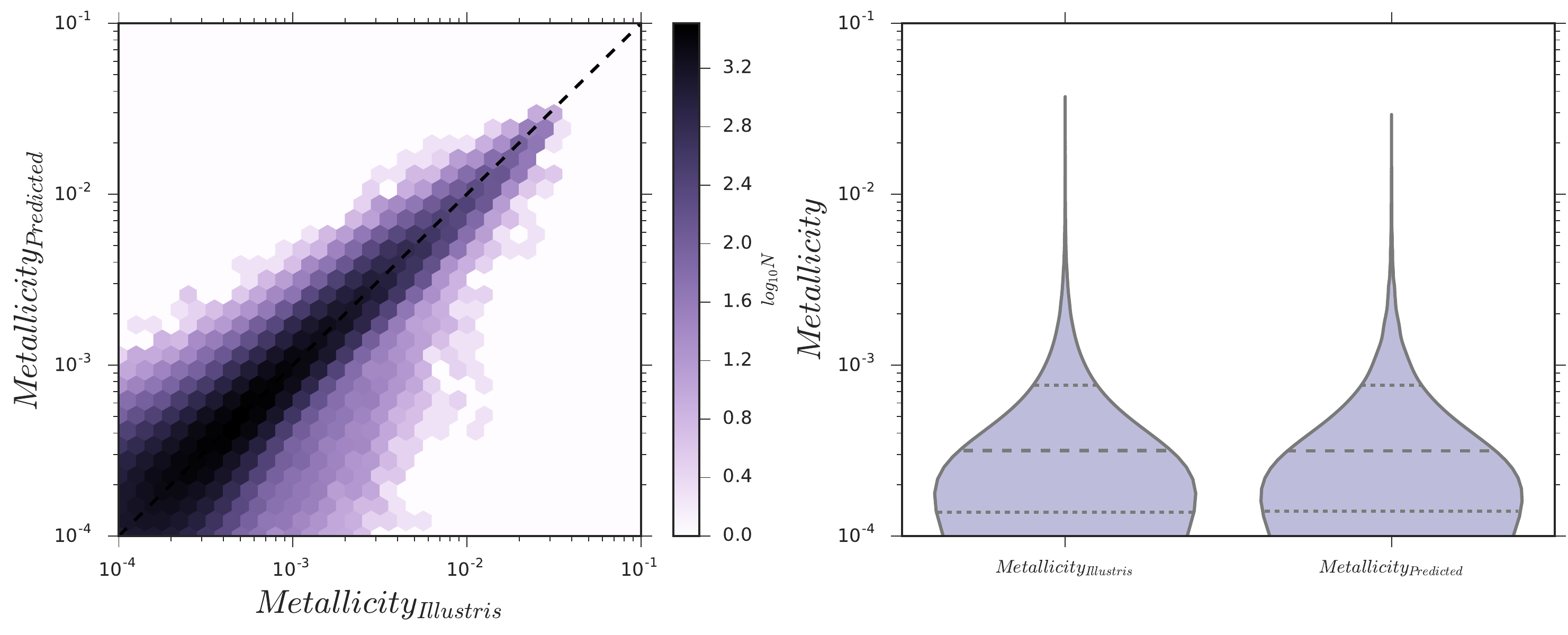}
    \caption{\textbf{\textit{Left}}: A hexbin plot of $Metallicity_{Illustris}$ and $Metallicity_{predicted}$ at $z=2$. The black dashed line corresponds to a perfect prediction. \textbf{\textit{Right}}: A violinplot showing the distributions of $Metallicity_{Illustris}$ and $Metallicity_{predicted}$. The median and the interquantile range are also shown.}
        \label{f10}
\end{figure*}

The stellar metallicity of an entire galaxy ($\frac{M_{z}}{M_{tot}}$) is also predicted at $z=0$ and $z=2$. The results for the stellar metallicity are shown in Table \ref{t6} and Figures \ref{f9} and \ref{f10}. Both plots indicate that the stellar metallicity is reconstructed very well using ML. There is some noticeable scatter at lower metallicities, similar to the other physical attributes that are discussed in this work. However, the distribution is reproduced very well implying that the predictions are statistically robust.

\par
\begin{figure*}
  \includegraphics[width=168mm]{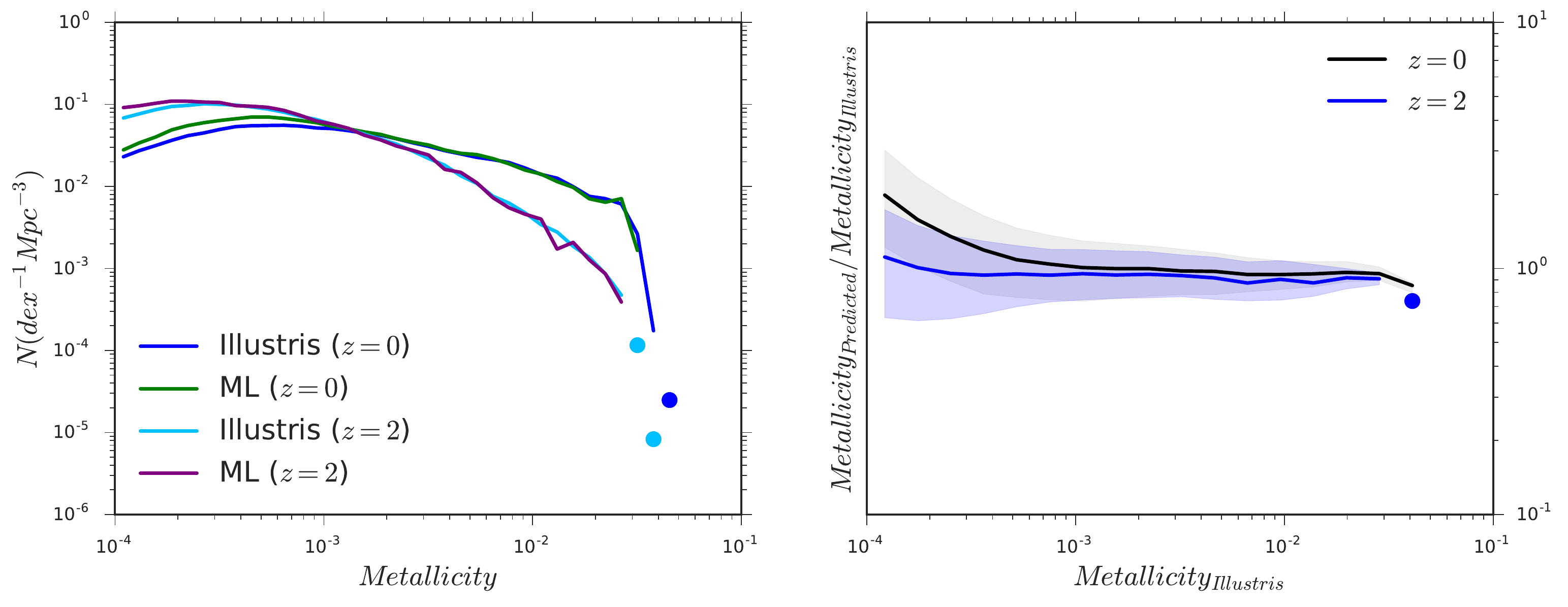}
\caption{\textbf{\textit{Left}}: The number density of galaxies as a function of stellar metallicity for the Illustris and the ML simulated galaxies at both $z=0$ and $z=2$. \textbf{\textit{Right}}: A binned plot of the fraction $\frac{Metallicity_{Illustris}}{Metallicity_{predicted}}$ as a function of $Metallicity_{Illustris}$. The area between the 25th and the 75th percentile for each bin has been shaded. Bins with less than 20 galaxies are denoted with big dots.}
        \label{f24}
\end{figure*}

 Furthermore, in Figure \ref{f24}, the number density of galaxies as a function of metallicity for the Illustris and the ML simulated galaxies at both $z=0$ and $z=2$ is shown alongside a binned plot of the fraction $\frac{Metallicity_{predicted}}{Metallicity_{Illustris}}$ as a function of $Metallicity_{Illustris}$. The number density  plot shows that the ML prediction at both epochs matches up very well with the Illustris galaxies. Furthermore, the fraction plot serves as a good object-by-object comparison of the ML and Illustris galaxies. The fraction plot shown in Figure \ref{f24} shows high scatter at lower values, which is consistent with what is shown in the hexbin plot, but shows an average value of $1$ for the ratio of the predicted metallicity to the Illustris metallicity. The statistically consistent results (from the hexbin plots) along with the mass function and the fraction plots imply that the population of ML simulated galaxies is consistent with the population of Illustris galaxies at both epochs.

\begin{table}
\caption{Metallicity statistics}
 \label{t6}
 \begin{tabular}{@{}lcccccc}
   Redshift & $MSE_b$ & $MSE$ & ($\frac{MSE_b}{MSE}$) & $\rho$
         & $R^2$\\
  \hline
   $z=0$ & $1.515 \times 10^{-5}$ & $9.22 \times 10^{-7}$ & 16.438  & 0.969 & 0.939\\
   \hline
   $z=2$ & $2.12 \times 10^{-6}$ & $2.10 \times 10^{-7}$ & 10.106 & 0.949 & 0.901\\
  \hline
 \end{tabular}
\end{table}

\subsubsection{Color}
\begin{figure*}
  \includegraphics[width=168mm]{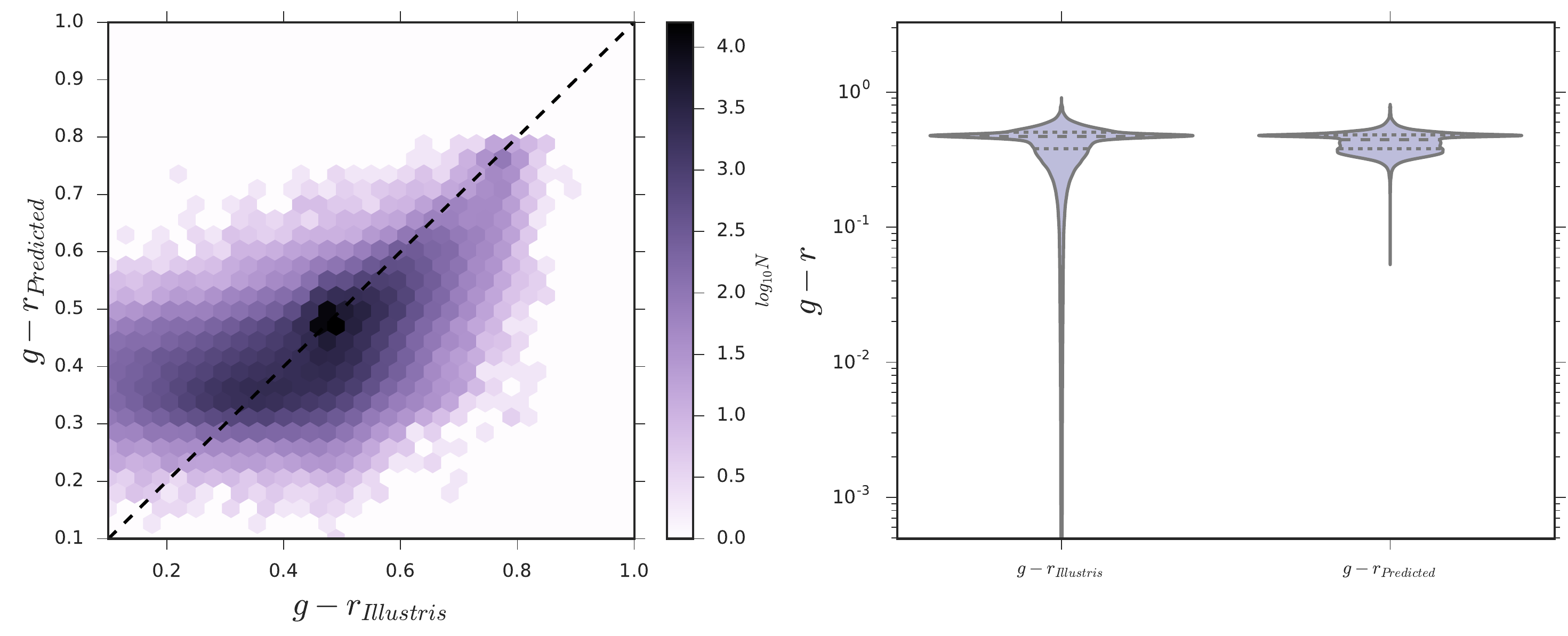}
    \caption{\textbf{\textit{Left}}: A hexbin plot of $(g-r)_{Illustris}$ and $(g-r)_{predicted}$ at $z=0$. The black dashed line corresponds to a perfect prediction. \textbf{\textit{Right}}: A violinplot showing the distributions of $(g-r)_{Illustris}$ and $(g-r)_{predicted}$. The median and the interquantile range are also shown.}
        \label{f11}
\end{figure*}

$g-r$ color is also predicted at $z=0$. The results are shown in Figure \ref{f11}. In Figure \ref{f11}, a hexbin and a violinplot plot are shown and in Figure \ref{f18}, $g-r$ is plotted as a function of stellar mass. The hexbin plot shows a lot of 
in the predicted values and a noticeable overprediction for lower magnitudes. However, the violinplot shows a peak at about the same point and the distributions look similar. This discrepancy may be because a few bins around $\approx 0.45$ have greater than $10^4$ galaxies in them.

\begin{figure*}
  \includegraphics[width=168mm]{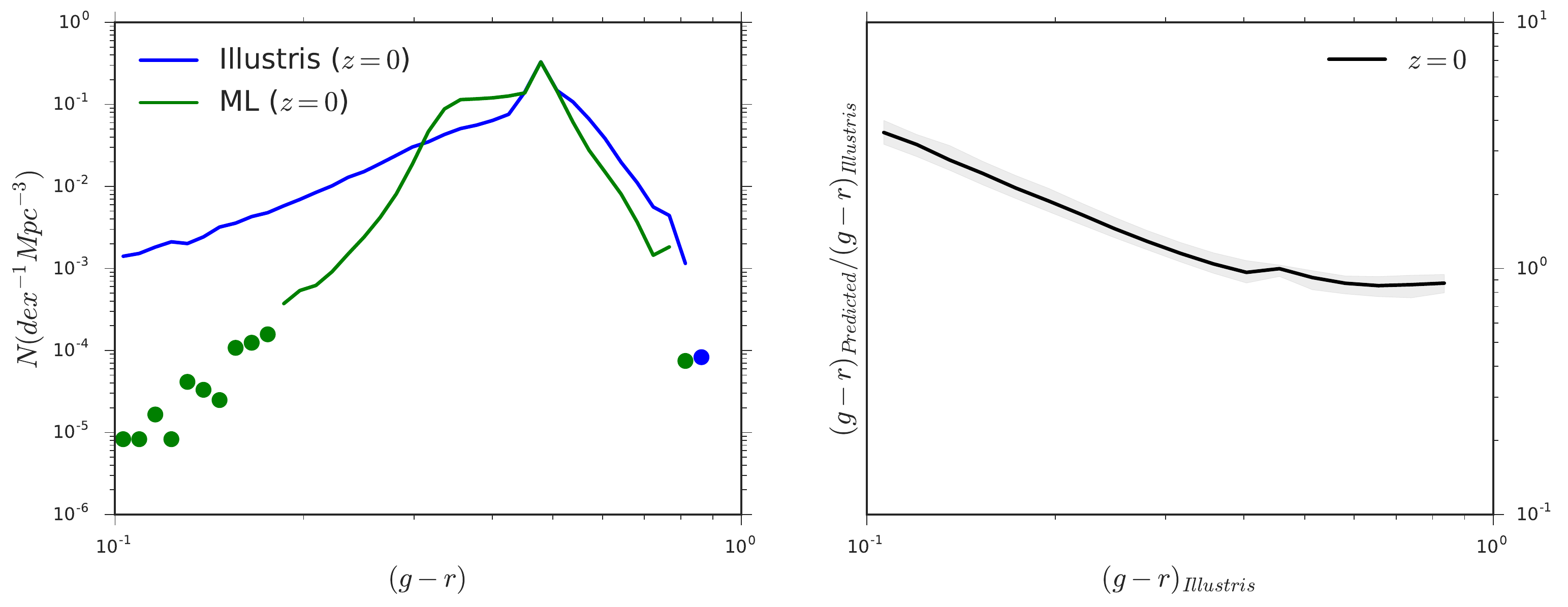}
\caption{\textbf{\textit{Left}}: The number density of galaxies as a function of $g-r$ for the Illustris and the ML simulated galaxies at both $z=0$ and $z=2$. \textbf{\textit{Right}}: A binned plot of the fraction $\frac{(g-r)_{Illustris}}{(g-r)_{predicted}}$ as a function of $(g-r)_{Illustris}$. The area between the 25th and the 75th percentile for each bin has been shaded. Bins with less than 20 galaxies are denoted with big dots.}
        \label{f22}
\end{figure*}

\par In Figure \ref{f22}, the number density of galaxies as a function of $g-r$ for the Illustris and the ML simulated galaxies at $z=0$ is shown alongside a binned plot of the fraction $\frac{(g-r)_{Illustris}}{(g-r)_{predicted}}$ as a function of $(g-r)_{Illustris}$. The number density  plot shows that the ML prediction at both epochs matches up reasonably well with the Illustris galaxies. Furthermore, the fraction plot serves as a good object-by-object comparison of the ML and Illustris galaxies. The fraction plot shown in Figure \ref{f22} shows low scatter at values and shows an average value of $3$ for  the ratio of $(g-r)_{predicted}$ to $(g-r)_{Illustris}$ at lower $g-r$ and approximately $1$ for $g-r$ $\geq$ 0.2. The statistically consistent results (from the hexbin plots) along with the mass function and the fraction plots imply that the population of ML simulated galaxies is consistent with the population of Illustris galaxies.

\par 
Figure \ref{f18} shows the $g-r$ color as a function of stellar mass. The Illustris and the predicted curves match up very well, placing confidence in our results. Following the recurring trend, the standard deviation in the predicted values is lower than that found in Illustris, i.e., there is more scatter with Illustris galaxies.

\begin{figure}
  \includegraphics[width=84mm]{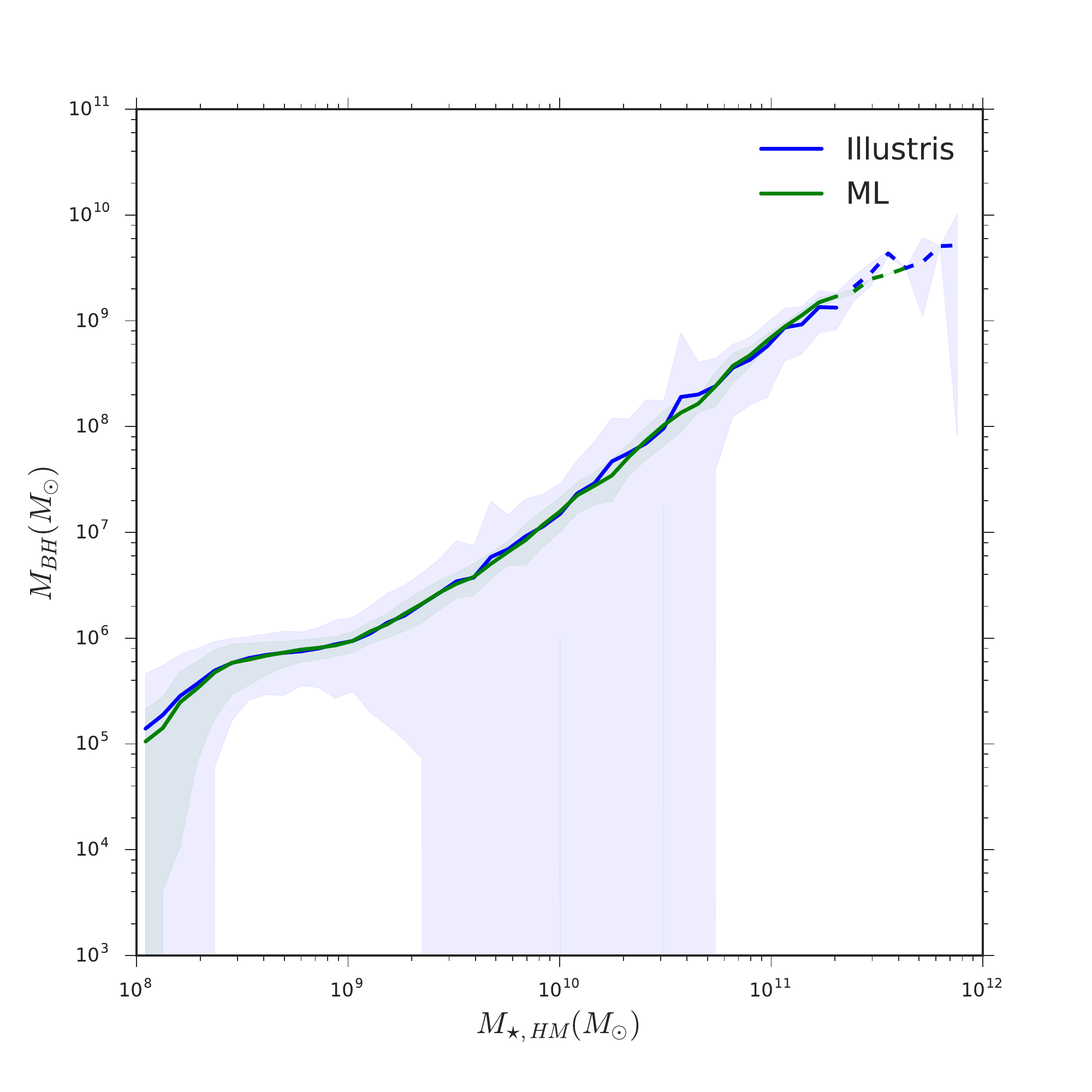}
    \caption{The BH-bulge mass relation at $z=0$ for the simulated ML galaxies and Illustris galaxies. Both quantities are binned using the stellar half mass. The two different shadings (blue for Illustris and green for ML) represent the standard deviation at each binned point.}
    \label{f12}

\end{figure}

\begin{figure}
  \includegraphics[width=84mm]{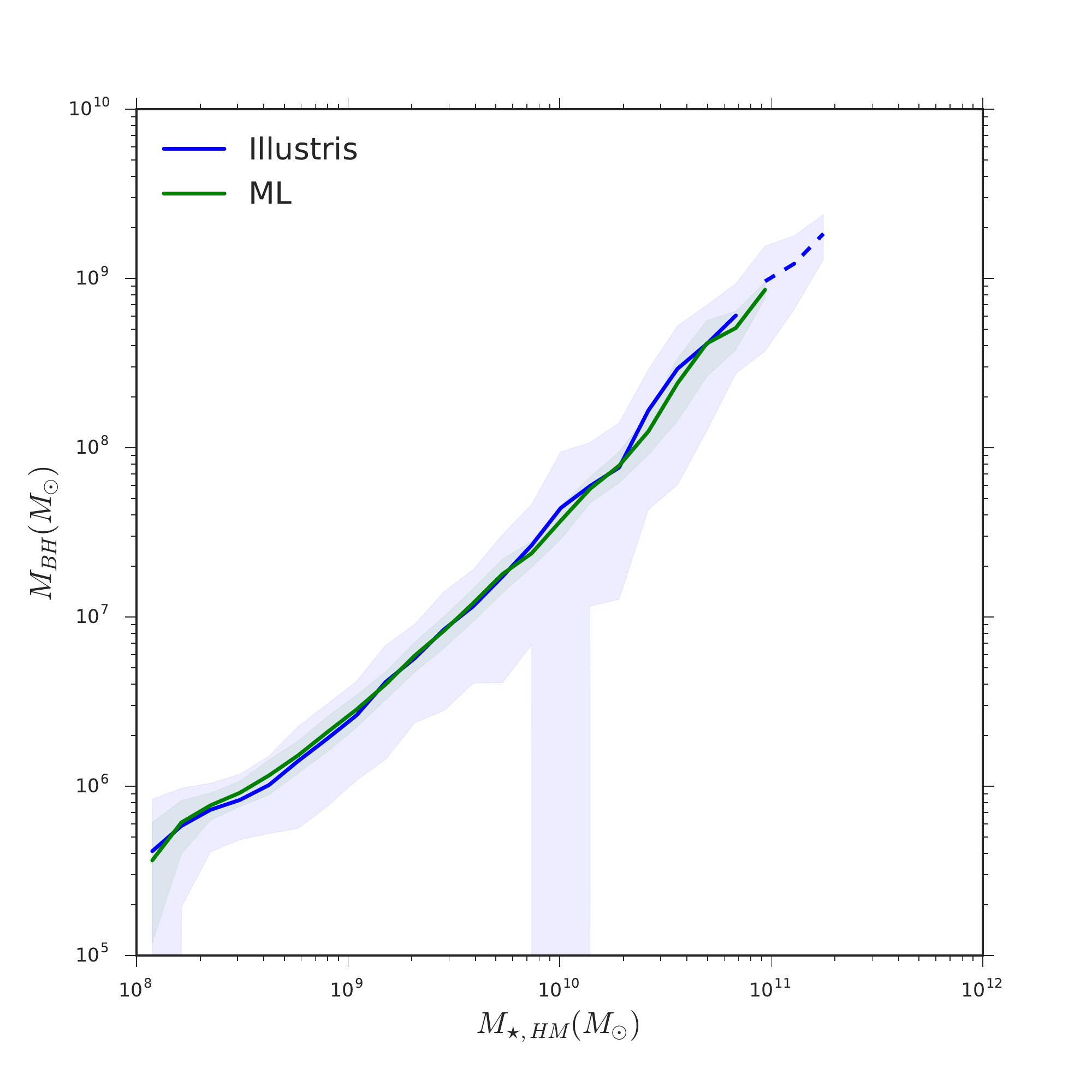}
    \caption{The BH-bulge mass relation at $z=2$ for the simulated ML galaxies and Illustris galaxies. Both quantities are binned using the stellar half mass. The two different shadings (blue for Illustris and green for ML) represent the standard deviation at each binned point.}
    \label{f13}    

\end{figure}

\begin{figure}
  \includegraphics[width=84mm]{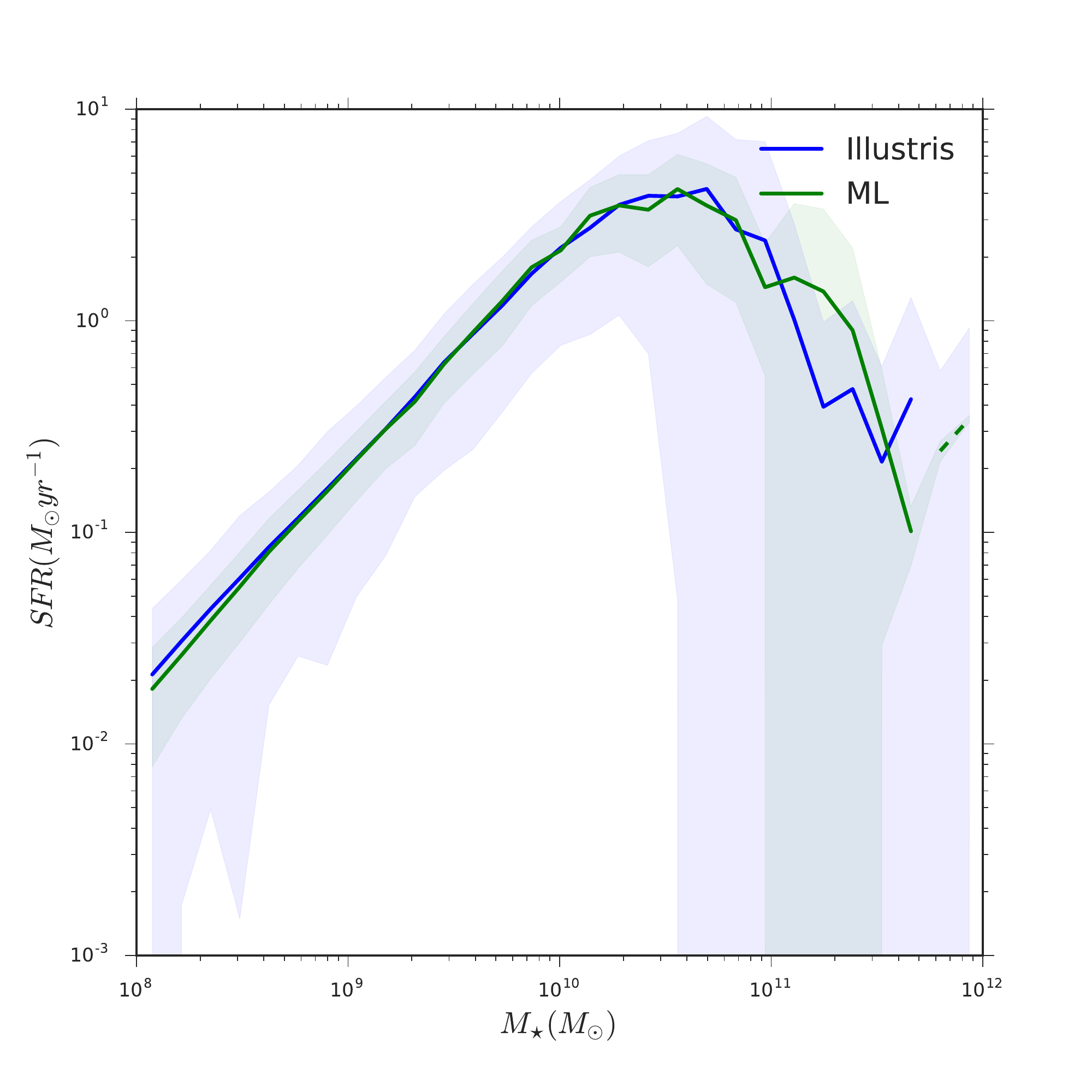}
        \caption{The SFR as a function of stellar mass at $z=0$ for the simulated ML galaxies and Illustris galaxies. Both quantities are binned using the stellar mass. The two different shadings (blue for Illustris and green for ML) represent the standard deviation at each binned point.}
    \label{f14}

\end{figure}
\begin{figure}
  \includegraphics[width=84mm]{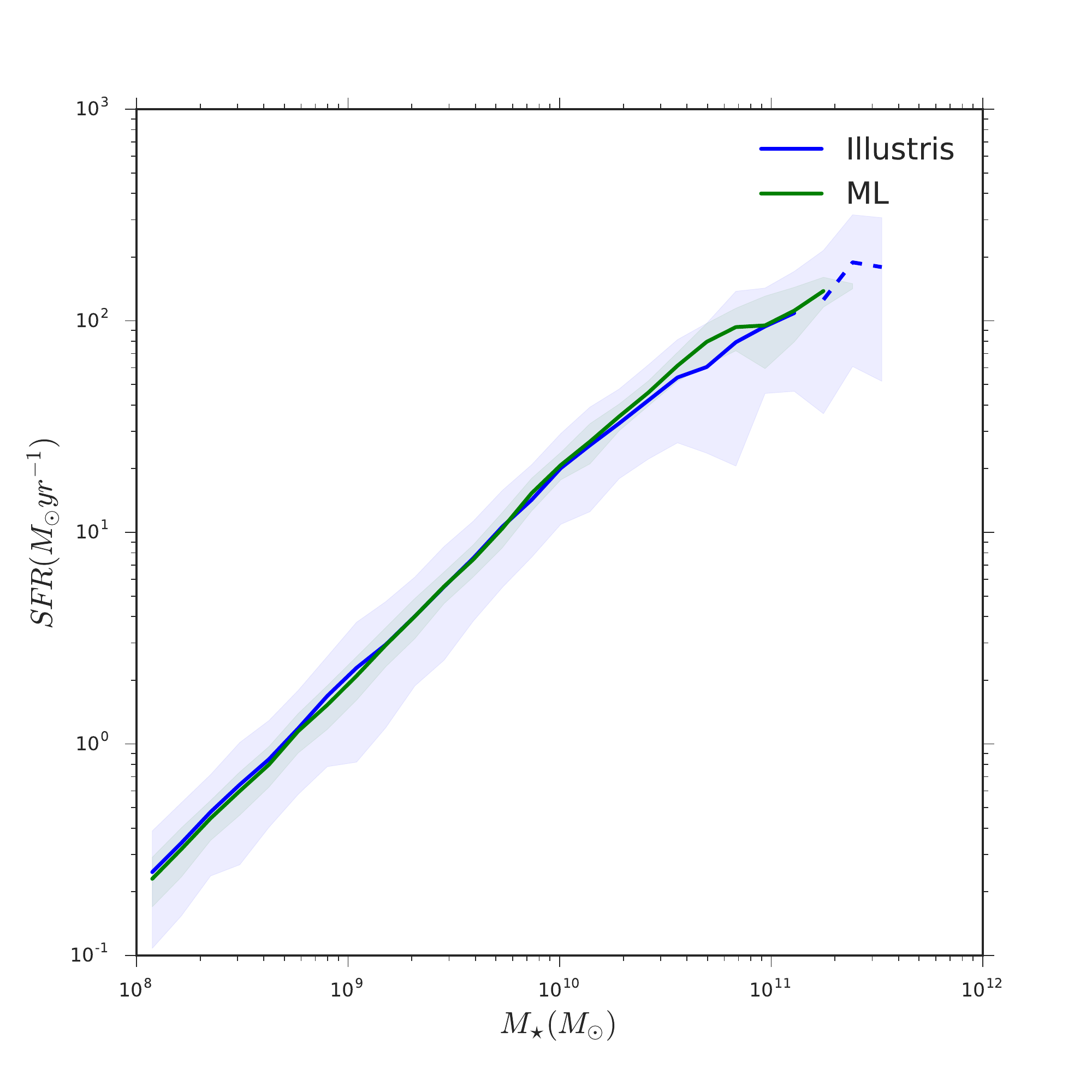}
    \caption{The SFR as a function of stellar mass at $z=2$ for the simulated ML galaxies and Illustris galaxies. Both quantities are binned using the stellar mass. The two different shadings (blue for Illustris and green for ML) represent the standard deviation at each binned point.}
    \label{f15}

\end{figure}

\begin{figure}
  \includegraphics[width=84mm]{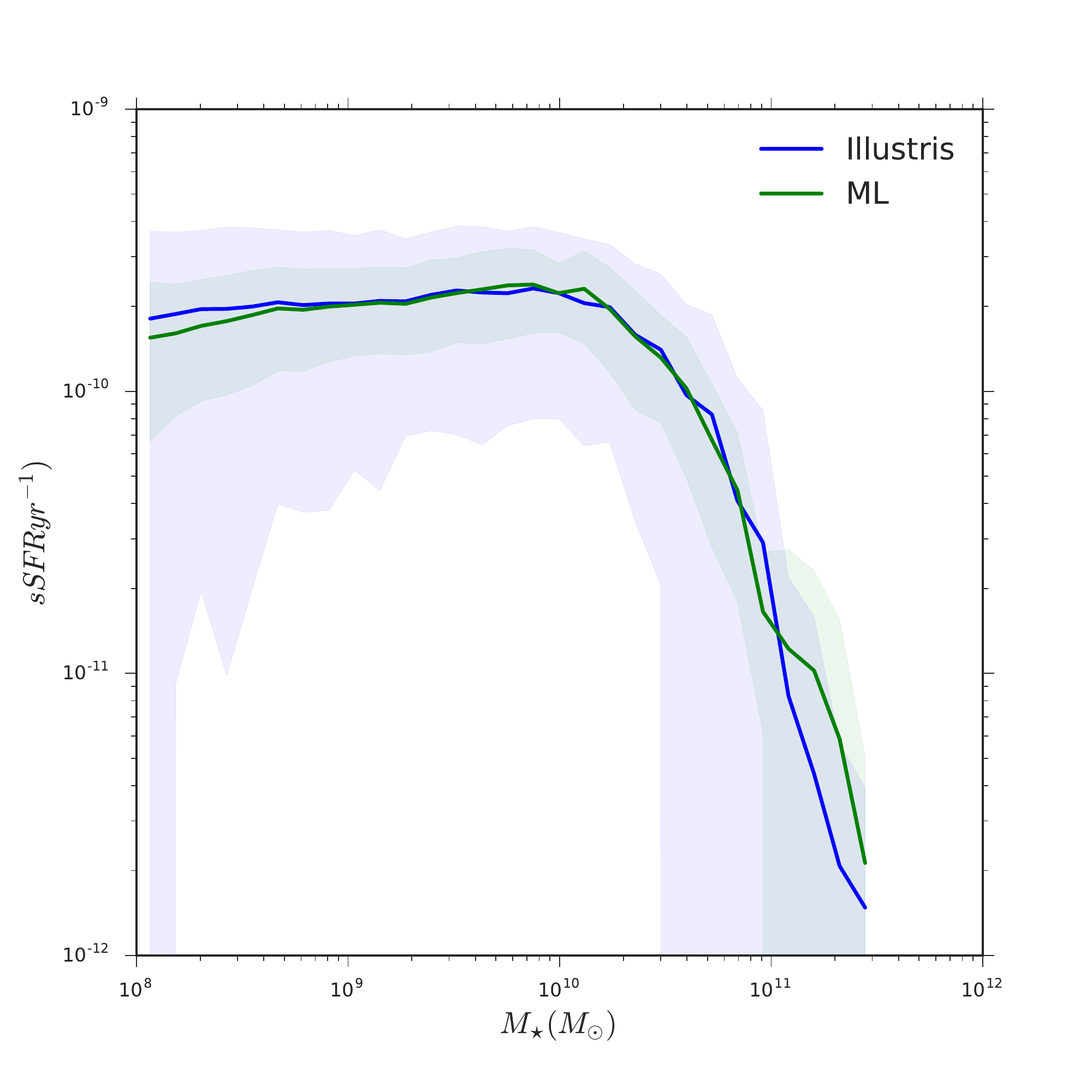}
    \caption{The SSFR as a function of stellar mass at $z=0$ for the simulated ML galaxies and Illustris galaxies. Both quantities are binned using the stellar mass. The two different shadings (blue for Illustris and green for ML) represent the standard deviation at each binned point.}
        \label{f16}
\end{figure}

\begin{figure}
  \includegraphics[width=84mm]{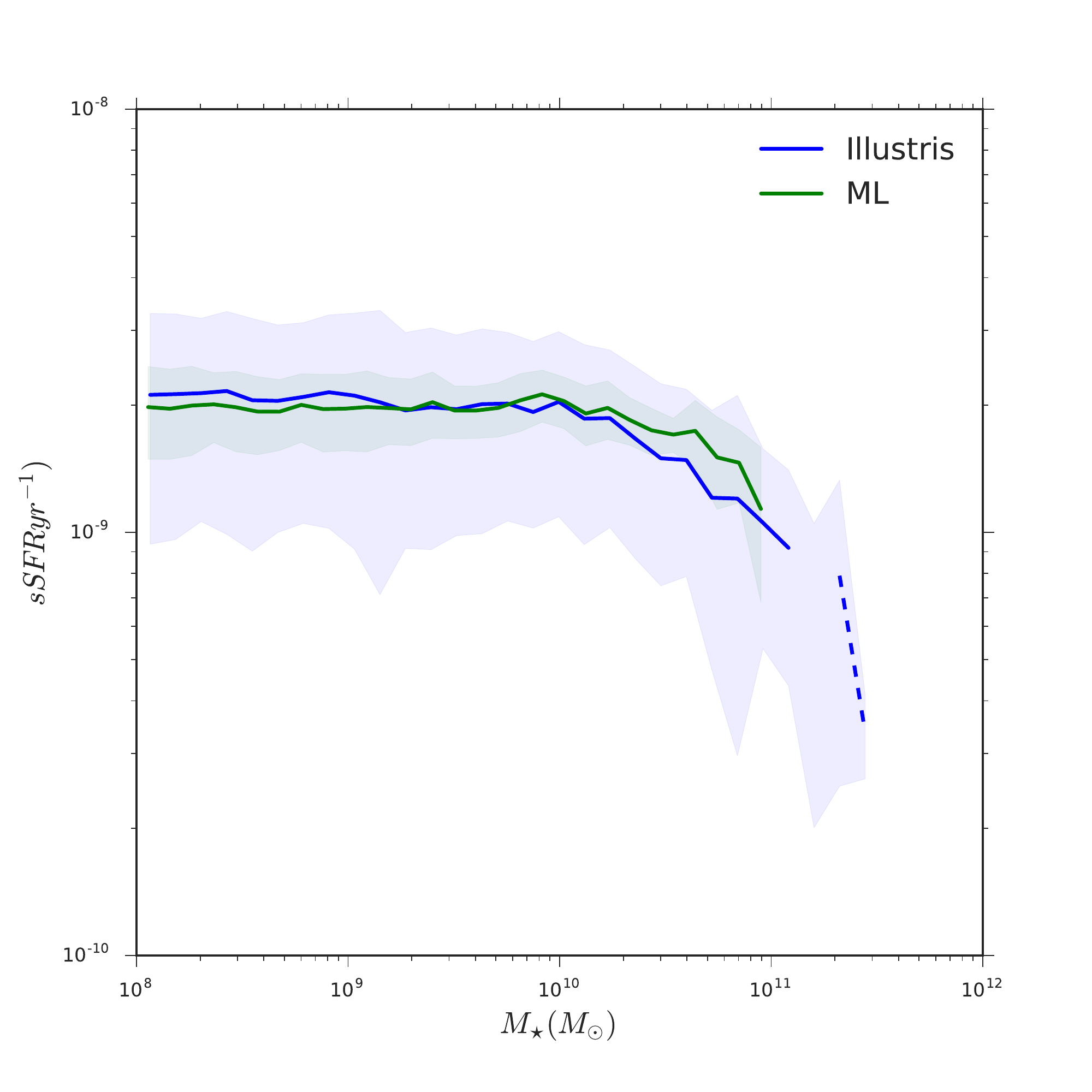}
    \caption{The SSFR as a function of stellar mass at $z=0$ for the simulated ML galaxies and Illustris galaxies. Both quantities are binned using the stellar mass. The two different shadings (blue for Illustris and green for ML) represent the standard deviation at each binned point.}
        \label{f17}
\end{figure}

\subsection{Dark Matter Halo Properties}
\par
Since we are using solely DM halo properties as inputs, it is worth exploring how much of a role each individual property plays in reproducing observations from hydrodynamical simulations. Previous studies \citep{neistein2011linking,jeeson2011correlation} have explored the importance of halo properties and correlations between the halo properties. 

\subsubsection{Feature Importance}
Here, we provide a feature importance plot that shows the relative importance of the halo properties in predicting the stellar mass at $z=0$. A more thorough analysis for predicting each individual galaxy property is planned for a future work.

\par
For tree-based machine learning techniques like ERT, the depth of a feature used as a decision node can be used to evaluate how important that particular feature is in the learning process. The expected fraction of the samples a feature contributes to can be used as an estimate of the relative importance of the features. We then average this quantity over all trees in the ensemble to get a less biased estimate for the importance of a particular feature.
\par
As shown in Figure \ref{f27}, the most important property in the prediction of the listed galaxy attributes is $V_{disp}$. The spin of the dark matter halo, on the other hand, plays a minimal role in the learning process. The feature importance helps us evaluate the impact that each DM property has on the prediction of the stellar mass. The results shown here are consistent with the feature selection shown below.

\subsubsection{Forward Feature Selection}

\par

We perform a simple two-step forward feature selection to explore what DM halo properties provide the tightest bound on the stellar mass. The first step of forward feature selection involves predicting the stellar mass using each individual DM feature. The MSE for each individual prediction is then calculated. In the second step, we select the feature that produced the lowest MSE and then individually add every other feature and repeat the analysis. The results are shown in table \ref{feature_selection}.

\begin{table}
\caption{Forward Feature Selection}
 \label{feature_selection}
 \begin{tabular}{@{}lcccccc}
   Property & First Round $MSE$ & Second Round $MSE$ \\
  \hline
    $M_{DM}$ & $0.2657$ & $0.1731$\\
   \hline
   $V_{disp}$ & $0.1599$ & $-$\\
  \hline
  $V_{max}$ & $0.1734$ & $0.1359$\\
  \hline
  $Spin_{x}$ & $1.1488 $ & $0.1898$\\
  \hline
  $Spin_{y}$ & $1.2999$ & $0.1697$\\
  \hline
  $Spin_{z}$ & $1.0555$ & $0.1766$\\
  \hline
  $N_{DM}$ & $0.2765$ & $0.1687$\\
  \hline
 \end{tabular}
\end{table}

The combined stellar mass MSE was $0.129$ using all attributes. The results show that $V_{disp}$ provides the tightest bound on $M_{\star}$, and $S_{x,y,z}$ does not contain as much information. In the second round, we can see that $V_{disp}$ together with $V_{max}$ provides an even tighter bound on $M_{\star}$. Overall, the most information about the stellar mass is contained in $V_{disp}$, $V_{max}$, $M_{DM}$, and $N_{DM}$. These results are consistent with what was found in the feature importance section above. The low MSE produced by just $V_{max}$ and $V_{disp}$ suggests that these inputs get pretty close to reproducing the lowest recorded MSE ($0.129$). The relatively simplistic forward feature selection study shows how many DM halo properties are needed to reproduce results from hydrodynamical simulations. A thorougher study that explores the effects on all galaxy attributes using a more sophisticated feature selection technique is a part of a future study. 

\section{Discussion} \label{disc}
The results presented above show that machine learning techniques are able to reproduce a strikingly similar population of galaxies to a full-blown hydrodynamical simulation by using only important physical properties of the dark matter halo in which the galaxy resides. The following physical attributes were predicted for each galaxy: gas mass, stellar mass, black hole mass, SFR, stellar metallicity, and $g-r$ color. There are two central differences between this work and the results presented in K16. First, we are using an N-body + hydrodynamical simulation where the baryonic physics employed is vastly more complicated and the treatment of physical processes is self-consistent. Second, the merger history was not included in our analyses reducing the computation time to a matter of minutes. 

\par
In the results shown previously, a wide variety of observed physical relationships are reproduced: the BH-bulge mass relation, the stellar mass-halo mass relation, SFR-stellar mass relation, SSFR-stellar mass relation, and color-stellar mass relation.  The fact that these important observational constraints are consistent with the Illustris galaxies is very promising. The reproduction of these relations is important because along with numerical and statistical robustness, the results show that the population of galaxies that is produced using ML is also physically consistent with the population of galaxies found in Illustris.

\par It is important to note here that our model is a purely phenomenological one. Unlike SHAMs, ML does not presuppose any relationship between the dark matter haloes and the galaxies residing in the haloes. ML, therefore, does not offer a replacement for SAMs or NBHS; instead, it can be used as a tool to explore the halo-galaxy connection and could be used as an analysis tool to explore how different simulation physics influences structure formation in the universe.

\par
An interesting note here is the similarity between our model and the methodology of SHAMs \citep{kravtsov2004dark, conroy2009connecting}. Both methods in question use physical halo properties to make statements about the properties of the galaxies that the haloes hold. A subtle, but important, distinction between the two methods must be made. SHAMs a priori assume that a relationship between the dark matter halo and the galaxy residing in the halo exists but ML does not make this assumption. Indeed, the reproduction of the SMHM places confidence in the key assumption that most SHAMs make: observable properties of galaxies are monotonically related to the dynamical properties of dark matter substructures. However, the ML technique is different from SHAMs in several ways. First and foremost, ML is able to recover a wider variety of galaxy properties. Second, ML, unlike SHAMs, does not assume any relation between any galaxy/halo property. ML is, essentially, going in "blind" and trying to learn the relationships that already exist. Furthermore, there is no fine-tuning involved. We would like to emphasize though that SHAMs provide a more prescriptive understanding (and are physically motivated) as compared to ML’s phenomenological nature and provide valuable physical insight that ML cannot. We posit that ML could prove to be a useful analysis tool in the context of the future study of hydrodynamical simulations because of its ability, by construction, to mimic hydrodynamical simulations.

\par 
The results obtained give a unique and powerful look into the halo-galaxy connection that is somewhat similar to that which we presented in K16. We are able to quantitatively and qualitatively show that there is a surprisingly strong environmental halo dependence for galaxy formation and evolution. The key difference between this work and K16 is that the results obtained herein show that the results obtained in K16 are valid in the context of N-body + hydrodynamical simulations. Furthermore, the computational costs associated with ML in the context of NBHS is miniscule. The full pipeline (preprocessing, running ML algorithms and plotting) took a total of 4 minutes for $z=0$ and 6 minutes for $z=2$, making ML 2-4 orders of magnitude faster than SAMs and 3-6 orders of magnitude faster than NBHS.

\par
\begin{figure}
  \includegraphics[width=84mm]{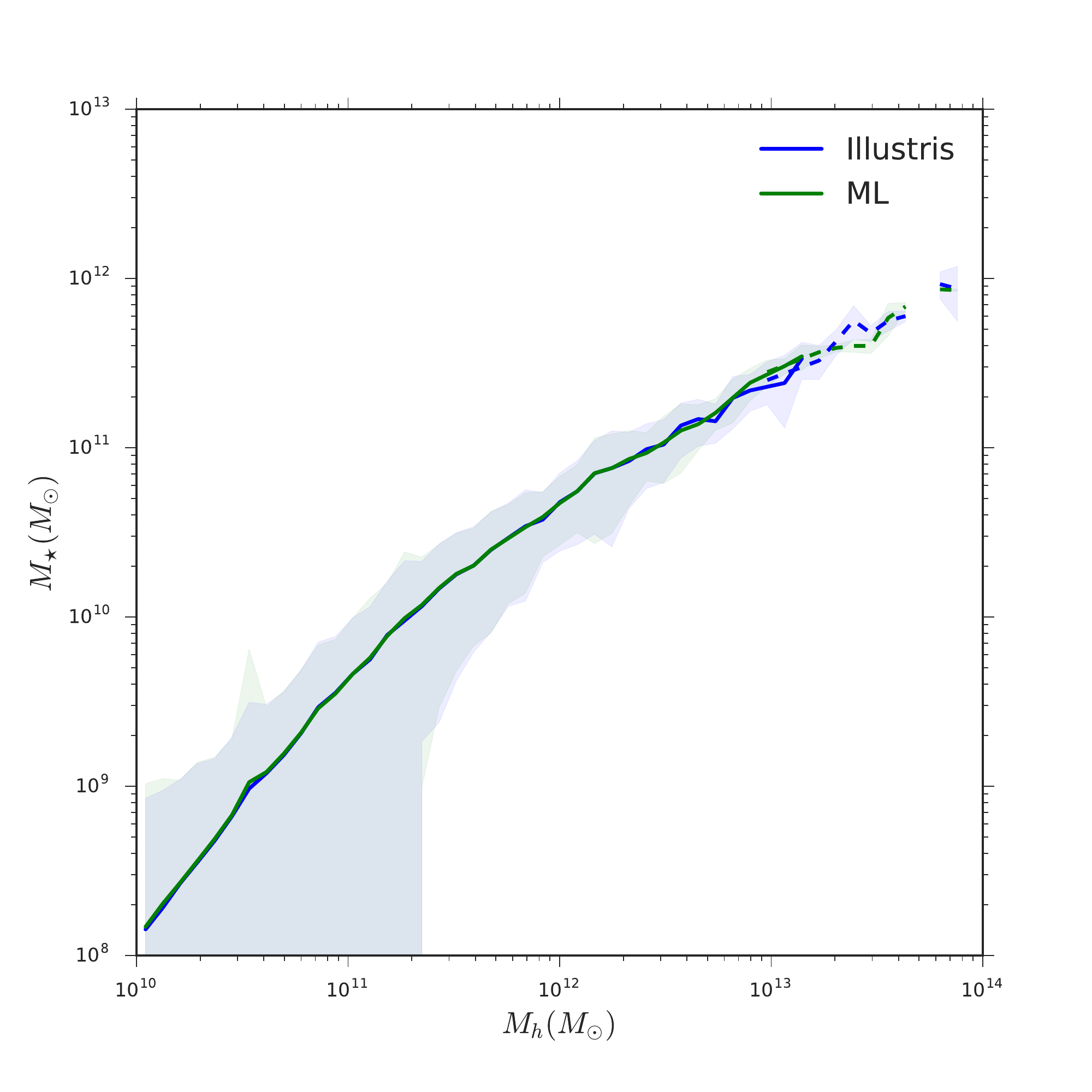}
    \caption{The stellar mass-halo mass relation at $z=0$ for the simulated ML galaxies and Illustris galaxies. Both quantities are binned using the halo mass. The two different shadings (blue for Illustris and green for ML) represent the standard deviation at each binned point.}
    \label{f18}
\end{figure}

\begin{figure}
  \includegraphics[width=84mm]{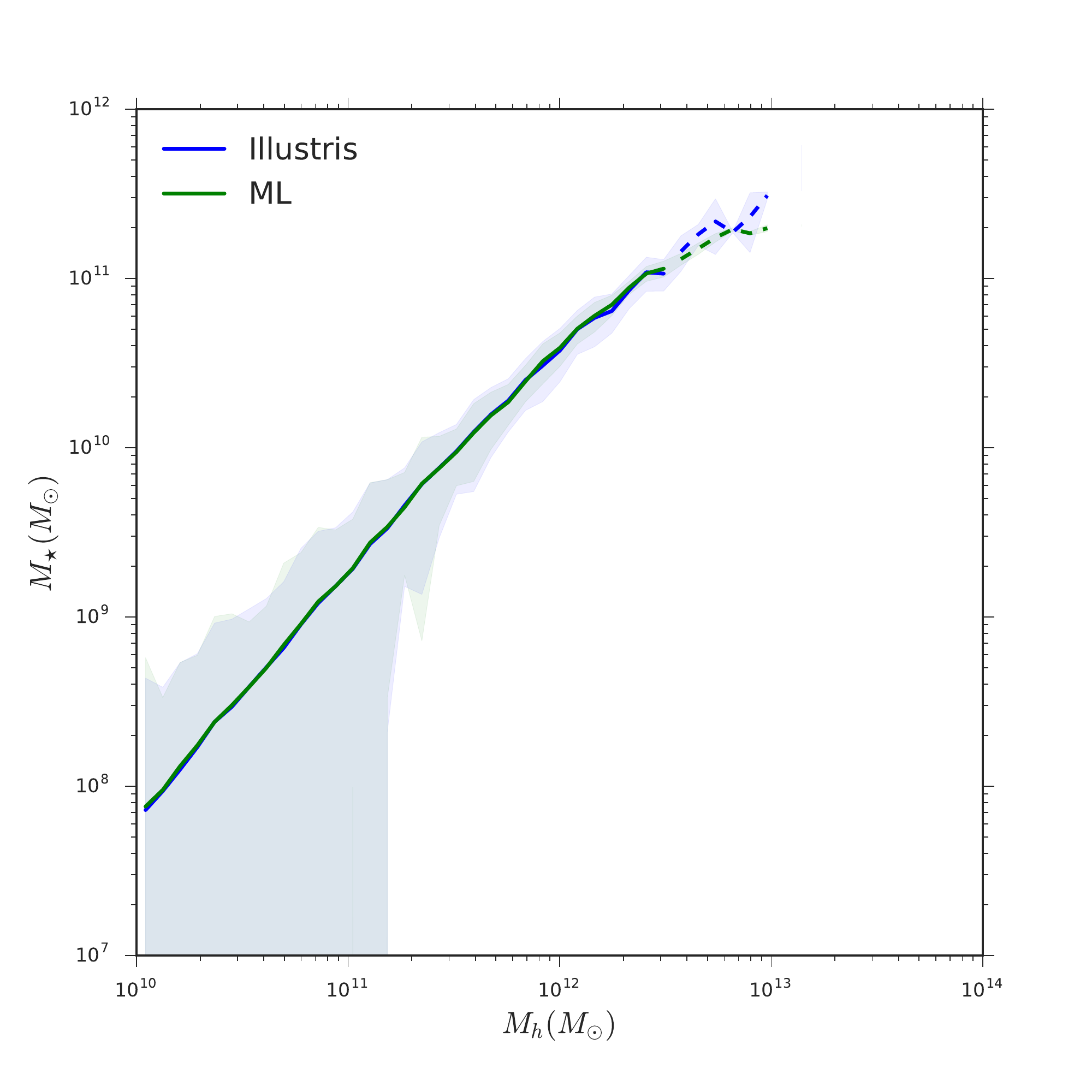}
    \caption{The stellar mass-halo mass relation at $z=2$ for the simulated ML galaxies and Illustris galaxies. Both quantities are binned using the halo mass. The two different shadings (blue for Illustris and green for ML) represent the standard deviation at each binned point.}
    \label{f19}
\end{figure}

 
A recurring discrepancy that we find between our results and the results presented in Illustris is the smaller scatter in our results (i.e. a lower standard deviation in ML galaxies compared to the Illustris galaxies). We believe that this is simply because of the lack of information. Machine learning is not able to fully learn the combined effects of the physical processes underlying the accumulation of some of the attributes simply because there just isn't enough information in the underlying dark matter properties about an attribute. For instance, as shown in Figure \ref{f18}, the scatter between Illustris and the ML galaxies is very similar. We would perhaps expect this because the relation between stellar mass and halo properties has been extensively studied and is the basis of SHAMs. However, the scatter between the ML and Illustris galaxies isn't as similar for, say, the color-stellar mass relation. We argue that this is because ML algorithms are able to obtain only a limited amount of information about the color from the DM properties and are unable to perfectly fit the relation. Therefore, given the scatter in ML for the SMHM and the robust predictions for stellar mass, we are inclined to believe that the scatter is indeed physical but incomplete in some cases due to a lack of information in the DM properties about baryonic evolution.  

\par
However, the goal of this work is not to exactly model each physical process with high accuracy and produce a numerically identical set of galaxies; instead, the goal is to explore the halo-galaxy connection in the backdrop of NBHS using a new technique and to evaluate how much information can be extracted from the dark matter properties about the eventual baryonic evolution of galaxies. The results shown in this work demonstrably show that a numerically, statistically, and physically robust population of galaxies is produced by ML when the algorithms are trained and tested on a robust N-body + hydrodynamical simulation. By physically robust, we mean that the combined effects of the physical processes that play a role in the evolution of one of the attributes above is at least approximately captured by ML algorithms. We believe that the ML simulated population of galaxies is physically robust for three reasons: statistical consistency with Illustris, the reproduction of various mass functions, and the positive results obtained from an object-by-object comparison. We note that there is noticeable scatter in the object-by-object comparison, especially at lower masses; however, the binned fraction plots demonstrably show that the ML recovered set of galaxies is, by and large, consistent with what is found in Illustris. The relative simplicity, the computational efficiency, and the physically consistent population of galaxies that is produced cements ML as an invaluable analysis tool in future galaxy formation studies. Our current approach does not replace any current galaxy formation models; instead, the applicability of ML lies in supplementing, validating, and extending current models. In a forthcoming work, ML is used to train algorithms on an NBHS and applied to completely independent N-body only simulations to populate the N-body only simulations with galaxies by, essentially, mimicking an NBHS. However, these mock galaxy catalogs are created in the order of minutes.

\begin{figure}
  \includegraphics[width=84mm]{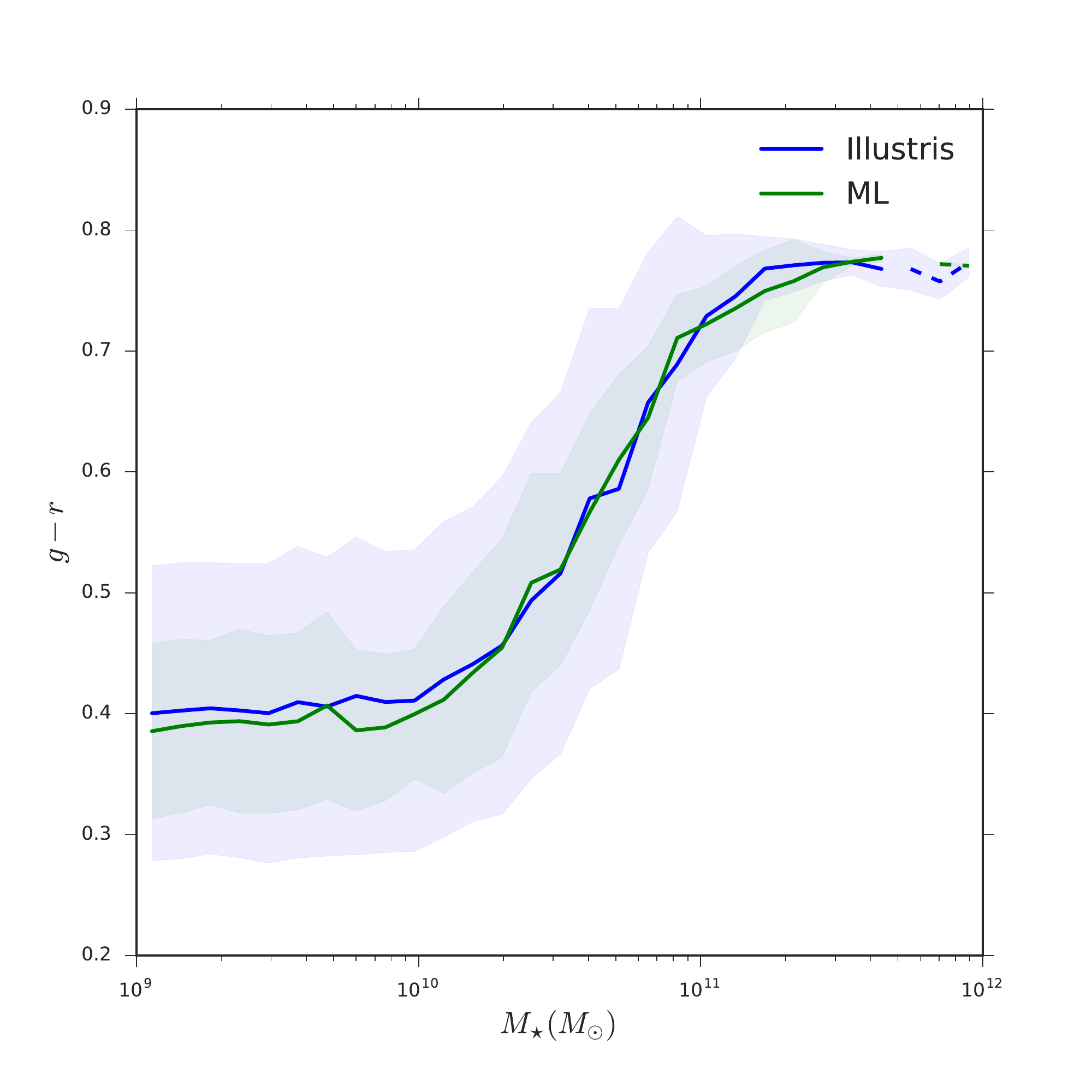}
    \caption{$g-r$ at $z=0$ as a function of stellar mass for the simulated ML galaxies and Illustris galaxies. Both quantities are binned using the stellar  mass. The two different shadings (blue for Illustris and green for ML) represent the standard deviation at each binned point.}
        \label{f20}

\end{figure}

\begin{figure}
\includegraphics[width=84mm]{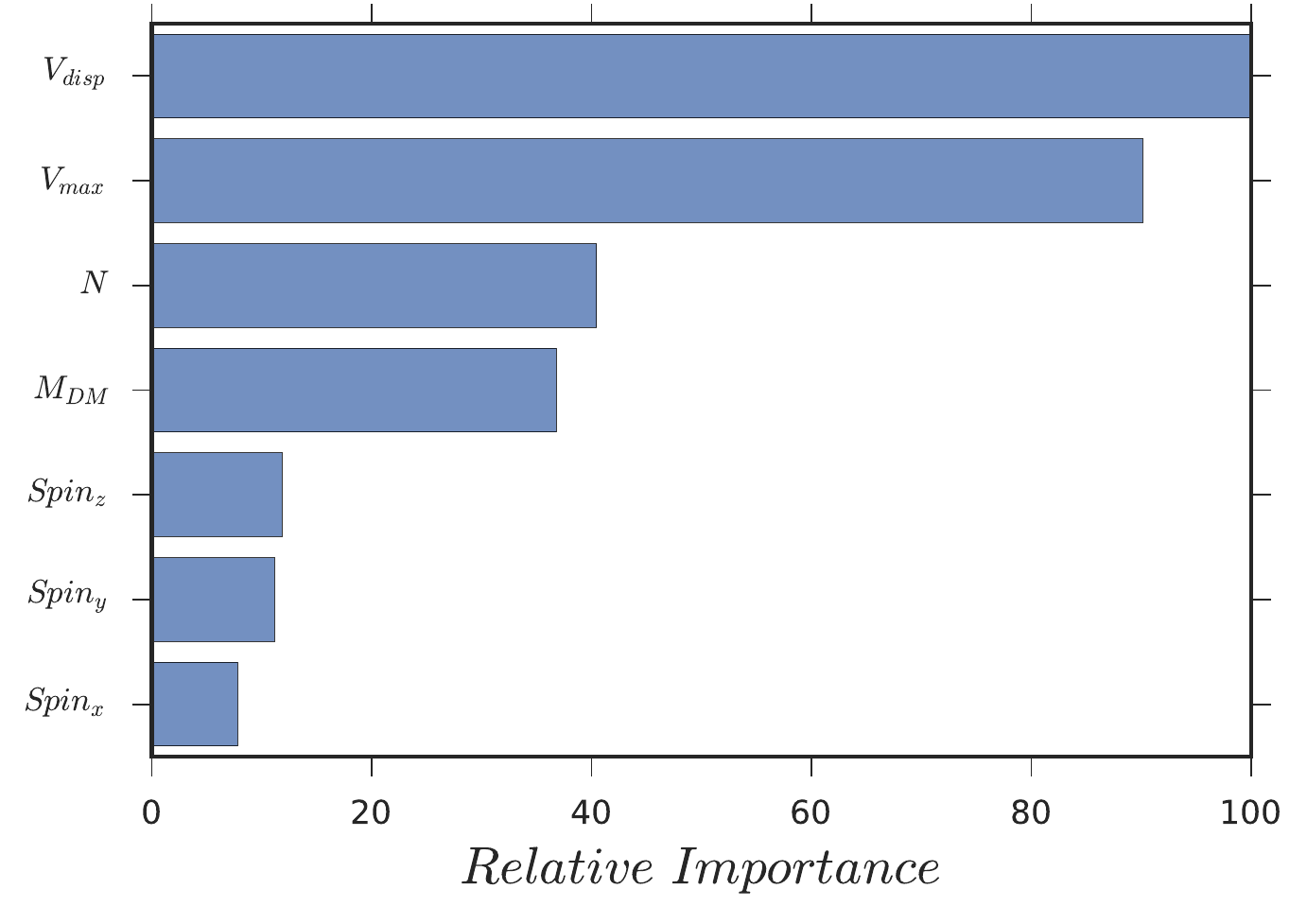}
\caption{The relative importance of different halo properties in predicting the stellar mass.}
\label{f27}
\end{figure}

\par
Another important point is the role of the subgrid models in N-body + hydrodynamical simulations. The subgrid models employed in Illustris are fairly similar to the prescriptions used in SAMs. However, there are more free parameters in SAMs that can be fine-tuned better but lead to degeneracies. Even though the method used to evolve physical quantities in hydrodynamical simulations is significantly more sophisticated, the same underlying physics helps explain why ML is successful at reproducing a physically reasonable population of galaxies. The success of ML in modeling galaxy formation is a statement on the ability of ML to infer these subgrid models and global observational constraints in the midst of hydrodynamical evolution.

\par
Overall, the results presented in this paper and our previous work on SAMs quantitatively show that an appreciable amount of information about galactic formation and evolution can be extracted from dark matter substructures. By reproducing some fundamental observational constraints, we show that ML is able to mimic a full-blown hydrodynamical simulation reasonably well, but 6 orders of magnitude faster. 

\section{Conclusions} \label{cl}
In this work, a variety of ML techniques were used to reconstruct a set of galaxies in an NBHS using solely the DM halo physical properties. Using the Illustris simulation to train and test the ML algorithms, the gas mass, stellar mass, BH mass, SFR, and $g-r$ are predicted. ML provides an incredibly unique and powerful framework for this particular problem for three reasons: simplicity of ML algorithms, computational efficiency of ML algorithms, and their ability to model incredibly complex relationships. The results shown in this work demonstrably show that a numerically, statistically and physically robust population of galaxies is produced by ML when the algorithms are trained and tested on an N-body + hydrodynamical simulation. 
\par 
Our primary conclusions are as follows:
\begin{enumerate}
 \renewcommand{\theenumi}{(\arabic{enumi})}
 \item
 Exploring the extent of the influence of dark matter haloes and its environment on galaxy formation and evolution is a non-trivial problem with poorly defined inputs and mappings. The problem is even murkier in the backdrop of hydrodynamical simulations, where the evolution is vastly more sophisticated and self-consistent with fewer fine-tuned parameters. ML offers a powerful framework to explore this problem. 
 \item 
 Using the Illustris simulation, a few important physical properties of dark matter haloes were used to predict the gas mass, stellar mass, black hole mass, SFR, stellar metallicity, and $g-r$ color. No baryonic processes were explicitly included in our analysis. We used extremely randomized trees for our analyses. 
 \item
 A remarkably similar population of galaxies is reconstructed when the ML algorithms are trained and tested on the Illustris simulation. The individual physical attributes are generally predicted quite well with few discrepancies. The ML simulated galaxies match up with the Illustris galaxies by following a variety of global constraints: the BH-bulge mass relation, the stellar mass-halo mass relation, SFR-stellar mass relation, SSFR-stellar mass relation, and color-stellar mass relation. 
 
 \item 
 A recurring, and important, discrepancy was the noticeably smaller scatter (lower standard deviation) for a given attribute in the ML galaxies compared to the Illustris galaxies. We believe that this is simply because of the lack of information. Machine learning is not able to fully learn the combined effects of the physical processes underlying the accumulation of some of the attributes simply because there just isn't enough information in the underlying dark matter properties about, say, the stellar metallicity. 
 
 \item
 However, the goal of this work was not to construct a population of galaxies that is numerically identical. Instead, the goal of this work goal was to evaluate how much information can be extracted from the dark matter only properties about the eventual baryonic evolution of galaxies. If the ML simulated galaxies are physically reasonable, which they are, then the approximate mapping found by ML between the dark matter halo properties and the galaxy properties solidifies ML's role in future galaxy formation studies. The conclusion is clear: ML is able to mimic how galaxies are evolved in an NBHS approximately well. Furthermore, ML is able to recreate the population of galaxies in about 4 minutes in contrast to millions of hours, prompting its effectiveness in future galaxy formation studies.
\end{enumerate}

\par 
The success of ML algorithms at modeling galaxy formation reasonably well in an NBHS opens up a wide variety of avenues for future work. Most notably, in a forthcoming work, we adopt this methodology to train an ML algorithm on Illustris (N-body + hydro) and apply it to three separate N-body only simulations (Illustris-dark, Bolshoi, and Dark Sky) to mimic an NBHS to populate an N-body only simulation with galaxies in the order of minutes. The approach laid out in this work could help with the rapid creation of mock galaxy catalogs for upcoming surveys. 

\section{Acknowledgments}
We would like to thank the referee for useful comments that improved the presentation of our results. HMK and RJB acknowledge support from the National Science Foundation Grant No. AST-1313415. HMK has been supported in part by funding from the LAS Honors Council at the University of Illinois and by the the Office of Student Financial Aid at the University of Illinois. HMK also thanks support from the Shodor Foundation and Blue Waters. RJB has been supported in part by the Center for Advanced Studies at the University of Illinois. MJT is supported by the Gordon and Betty Moore Foundation's Data-Driven Discovery Initiative through Grant GBMF4561.

The Illustris-1 simulation was run on the CURIE supercomputer at CEA/France as part of PRACE project RA0844, and the SuperMUC computer at the Leibniz Computing Centre, Germany, as part of GCS-project pr85je. The further simulations were run on the Harvard Odyssey and CfA/ITC clusters, the Ranger and Stampede supercomputers at the Texas Advanced Computing Center through XSEDE, and the Kraken supercomputer at Oak Rridge National Laboratory through XSEDE.
\footnotesize{
	\bibliographystyle{mnras}
	\bibliography{main}
}

\bsp
\label{lastpage}
\end{document}